\pdfoutput=1
\documentclass[fleqn,usenatbib]{mnras}

\usepackage[T1]{fontenc}
\usepackage{ae,aecompl}

\usepackage{graphicx}	
\usepackage{amsmath}	
\usepackage{amssymb}	
\usepackage [applemac]{inputenc}	
\usepackage{subfig}

\title[Gas kinematics in the inner kiloparsec of NGC 1365]{Ionized gas kinematics within the inner kiloparsec of the Seyfert galaxy NGC 1365}

\author[D. Lena et al.]{Davide Lena,$^{1,2,3}$\thanks{E-mail: D.Lena@sron.nl}
Andrew Robinson,$^{1}$
Thaisa Storchi-Bergmann,$^{4}$\newauthor
Guilherme S. Couto,$^{4}$
Allan Schnorr-M\"{u}ller,$^{5,6}$ \&
Rogemar A. Riffel$^{7}$
\\
$^{1}$School of Physics and Astronomy, Rochester Institute of Technology, 84 Lomb Memorial Drive, Rochester, NY 14623-5603, USA\\
$^{2}$SRON, Netherlands Institute for Space Research, Sorbonnelaan 2, NL-3584 CA Utrecht, The Netherlands\\
$^{3}$Department of Astrophysics/IMAPP, Radboud University, Nijmegen, PO Box 9010, NL-6500 GL Nijmegen, The Netherlands\\
$^{4}$Instituto de F\'{i}sica, Universidade Federal do Rio Grande do Sul, 91501-970, Porto Alegre, Brazil\\
$^{5}$Max-Planck-Institut f\"{u}r extraterrestrische Physik, Giessenbachstr. 1, D-85741, Garching, Germany\\
$^{6}$CAPES Foundation, Ministry of Education of Brazil, 70040-020, Brasília, Brazil\\
$^{7}$Universidade Federal de Santa Maria, Departamento de F\'{i}sica, 97105-900, Santa Maria, RS, Brazil
}

\date{Accepted for publication on MNRAS}

\pubyear{2015}

\begin{document}
\label{firstpage}
\pagerange{\pageref{firstpage}--\pageref{lastpage}}
\maketitle

\begin{abstract}
We observed the nuclear region of the galaxy NGC 1365 with the integral field unit of the Gemini Multi Object Spectrograph mounted on the GEMINI-South telescope. The field of view covers $13^{\prime\prime} \times 6^{\prime\prime}$ ($1173 \times 541$ pc$^{2}$) centered on the nucleus, at a spatial resolution of $52$ pc. The spectral coverage extends from $5600$ \AA\ to $7000$ \AA, at a spectral resolution $R=1918$. NGC 1365 hosts a Seyfert 1.8 nucleus, and exhibits a prominent bar extending out to $100^{\prime\prime}$ (9 kpc) from the nucleus. The field of view lies within the inner Lindblad resonance. Within this region, we found that the kinematics of the ionized gas (as traced by [OI], [NII], H$\alpha$, and [SII]) is consistent with rotation in the large-scale plane of the galaxy. While rotation dominates the kinematics, there is also evidence for a fan-shaped outflow, as found in other studies based on the [OIII] emission lines. 
Although evidence for gas inflowing along nuclear spirals has been found in a few barred galaxies, we find no obvious signs of such features in the inner kiloparsec of NGC 1365. However, the emission lines exhibit a puzzling asymmetry that could originate from gas which is slower than the gas responsible for the bulk of the narrow-line emission. We speculate that it could be tracing gas which lost angular momentum, and is slowly migrating from the inner Lindblad resonance towards the nucleus of the galaxy.
\end{abstract}




\section{Introduction}

The powerful non-stellar radiation observed in active galactic nuclei (AGNs) originates from gas accretion onto a supermassive black hole \citep[SMBH, e.g.][]{Lynden-Bell69,Soltan82}. Accretion rates as low as 10$^{-3}$ M$_{\odot}$ yr$^{-1}$ are sufficient to reproduce the luminosity of local AGNs, however, despite the large amounts of gas present in galaxies, only 43\% of them are active \citep{Ho2008}. Gas transfer from kpc scales down to sub-pc distances from the SMBH is, at the time of writing, a lively research topic. Reviews of the subject have been presented by \citet{ShlosmanBF90,Knapen05,Jogee06,Alexander2012}.  

Theoretical work \citep[e.g.][]{Athanassoula92,HopkinsQ10,EmsellemRB15} and observations \citep[e.g.][]{GarciaBuCS05} show that different mechanisms are efficient in the removal of gas angular momentum at different scales. Torques due to kpc-scale bars, or to interactions and mergers, are efficient at channelling gas from several kpc down to the inner few 100 pc. When a bar is present, gas stalls and accumulates at the inner Lindblad resonance resulting in rings of gas \citep[e.g.][]{JogeeBS01} or star formation regions \citep[e.g.][]{PerezRamKP2000}. Gas might then be funneled within a few parsecs from the SMBH, or even closer, by nuclear bars, nuclear spirals \citep[e.g.][]{HopkinsQ10}, or because of dynamical friction \citep{JogeeKS99} and viscosity \citep{Bekki2000}.

When a reservoir of gas is present in the nuclear region of an active galaxy, then an outflow may take place \citep[e.g.][]{CrenshawKG03MassLoss}. Simulations suggest that AGN feedback has a crucial role in the evolution of the host galaxy \citep[e.g.][]{DiMatteoSH05}, however the modeling of the AGN physics is still simplistic and robust observational constraints are still far from being set.

Kinematical features consistent with gas inflows and/or outflows on scales of 10 - 100 pc have been observed in a number of nearby galaxies \citep[e.g.][]{FathiSBREtAl06,SBergmannEtAl07,RiffelEtAl08,MullerSanchezDG09,SMullerEtAl11,GarciaBuC12,CombesGBC14,DaviesMH14,SMSBN14,SchnorrMSBNF14,LenaRSB14,SmajiME15,ScharwachterDS15,LuoHB16}. However, a clear picture of the chain of events which ultimately feeds the AGN is yet to be established. How much of the accreted rest-mass energy is transferred to the interstellar medium through the outflows is still in the process of being determined.

Here we present a study of the ionized gas kinematics in the nuclear region of NGC 1365, one of the most extensively observed galaxies in the southern hemisphere (see \citealt{Lindblad99} for a detailed review of the early work). This is part of an ongoing study aimed at identifying correlations between the circum-nuclear gas kinematics and the SMBH accretion rate.

\vskip5pt
NGC 1365 is an archetypal barred galaxy, classified as SB(s)b by \cite{deVac91}. \citet{JonesJ80} identify the galaxy as a member of the Fornax cluster. It is known to host a Seyfert-like AGN which is classified as Seyfert 1.8 by \citet{VeronCV2006}. The nuclear region exhibits signatures of both star-formation and an AGN.
Evidence that star formation plays a major role in the nucleus of the galaxy was, perhaps, first identified by \citet{Morgan58} who reported the presence of ``hot spots", i.e. bright HII regions. More recently, the coexistence of AGN and stellar photoionization, in the vicinity of the nucleus, has been nicely mapped with integral field observations by \citet{SharpB10}.

Signs of nuclear activity have been identified mainly at X-ray and optical wavelengths: \citet{IyomotoMF97} observe a strong FeK emission line, and a point-like hard X-ray source consistent, in position, with the optical nucleus; a power-law component was judged necessary to perform a satisfactory fit to nuclear X-ray spectra in \citet{KomossaS98}. \citet{RisalitiEF05} report on extreme X-ray variability of the hard X-ray continuum. This was later interpreted as the result of variations in the distribution of discrete absorbers along the line of sight \citep[e.g.][and references therein]{BraitoRG14}. Such findings lend support to the hypothesis that the hard X-ray emission in the nucleus of NGC 1365 is a genuine AGN signature. The recent study of \citet{NardiniGR15}, which is based on data from the Chandra High Energy Transmission Grating, reveals a rich spectrum of soft X-ray emission lines from photoionized gas in the immediate vicinity of the AGN. They show evidence of line broadening, outflow, and possibly inflows, with velocities of the order of $1000$ km s$^{-1}$.

Optical spectroscopy revealed the presence of broad Balmer emission lines, narrow-line ratios typical of AGNs \citep[e.g.][]{VeronLZV80,EdmundsP82,SchulzKSM99}, and the presence of [NeV] and HeII \citep{PhillipsF80} indicating photoionization by a hard extreme-UV continuum, as is characteristic of AGNs. Fabry-Perot interferometry and narrow-band imaging showed a fan-shaped [OIII]$\lambda5007$ emission region extending about 10$^{\prime\prime}$ ($\approx 1$ kpc) south-eastward of the nucleus \citep{EdmundsTT88,StorchiBergmannB91,KristenJLB97}. 

\citet{AlonsoHerreroSR12} present an infrared study of the inner 5 kpc. They infer that most of the star formation within the inner Lindblad resonance (r $\approx 30^{\prime\prime}$, or $2.7$ kpc, \citealt{LindbladLA96}) takes place within obscured regions, in a nuclear star-forming ring of radius $\approx 1$ kpc. Furthermore, they propose that the photometric and spectral emission can be reproduced with a model including a torus with outer radius $r_{o} \approx 5$ pc, opening angle $\sigma \approx 36^{\circ}$, and with an AGN bolometric luminosity of $L_{bol} \approx 2.6 \times 10^{43}$ erg s$^{-1}$.

Evidence of AGN activity at radio wavelengths is scarce: \cite{StevensFN99} consider evidence for a radio jet to be marginal at best, suggesting that star formation dominates the energetics at radio, optical, and soft X-ray wavelengths.

Being a nearby galaxy, isolated, with a prominent bar and spiral arms, and having an intermediate inclination with respect to the observer's line of sight, the gas kinematics in NGC 1365 have been extensively studied. This line of investigation was begun by \citet{BurbidgeB60} and \citet{BurbidgeBP62} who used long slit spectroscopy to derive the rotation curve from H$\alpha$ and [NII] observations. Significantly different kinematics were derived by \citet{PhillipsEPT83} from observations of higher excitation lines, namely [OIII] and HeII. \citet{JorsaterM95} used high resolution observations of neutral hydrogen to map the large-scale velocity field. They determined the inner disk ($120^{\prime\prime} \lesssim r \lesssim 240^{\prime\prime}$) inclination to be 40$^{\circ}$. They also found non-circular motions associated with the bar, and a central reservoir of molecular gas. In the central region of the galaxy, a strongly disturbed velocity field was also revealed with the aid of optical observations by \citet{LindbladHHJ96} while a biconical outflow model was proposed by \citet{HjelmL96} to explain the [OIII] velocity field.

More recently, measurements of the large-scale 2D velocity field of H$\alpha$, and a new analysis of archival HI data were presented in \citet{ZanmarSSW08} who found strong asymmetries in the distribution of gas, dust and kinematical features, in both H$\alpha$ and HI. 

The galaxy NGC 1365 shares a number of similarities with NGC 1097, including morphology, distance, and inclination. It is not by chance that a number of authors have studied the two galaxies in parallel \citep[e.g.][]{BurbidgeB60,OndrechenH89,OndrechenHH89,BeckFS05}. In an earlier study of NGC 1097, \citet{FathiSBREtAl06}, members of our team identified gas inflows along a nuclear spiral which leads down to a few parsecs from the unresolved AGN. It seems natural to look for similar inflows in NGC 1365. 
While previous spectroscopic studies have focused on the large-scale kinematics of NGC 1365, here we present optical integral field spectroscopy for the inner 6\arcsec. We adopt the distance of 18.6 $\pm$ 0.6 Mpc, as determined from Cepheid variables by \citet{MadoreFS99}, which results in the scale $s =$ 90.2 pc/arcsec. 
\vskip5pt

The paper is organized as follows: observations and data reduction are described in \textsection \ref{sec: analysis}, the data analysis, with details on the emission lines fitting procedures, is presented in \textsection \ref{sec: fitting}. Results are given in \textsection \ref{sec: results} and discussed in \textsection \ref{sec: disc}. A summary of our findings is given in \textsection \ref{sec: conclusion_NGC1365} . 

\begin{figure*}
 \begin{minipage}[t]{.49\linewidth}\vspace{0pt}
\includegraphics[trim=3cm 1.2cm 4.825cm 0.525cm, clip=true, scale=0.8]{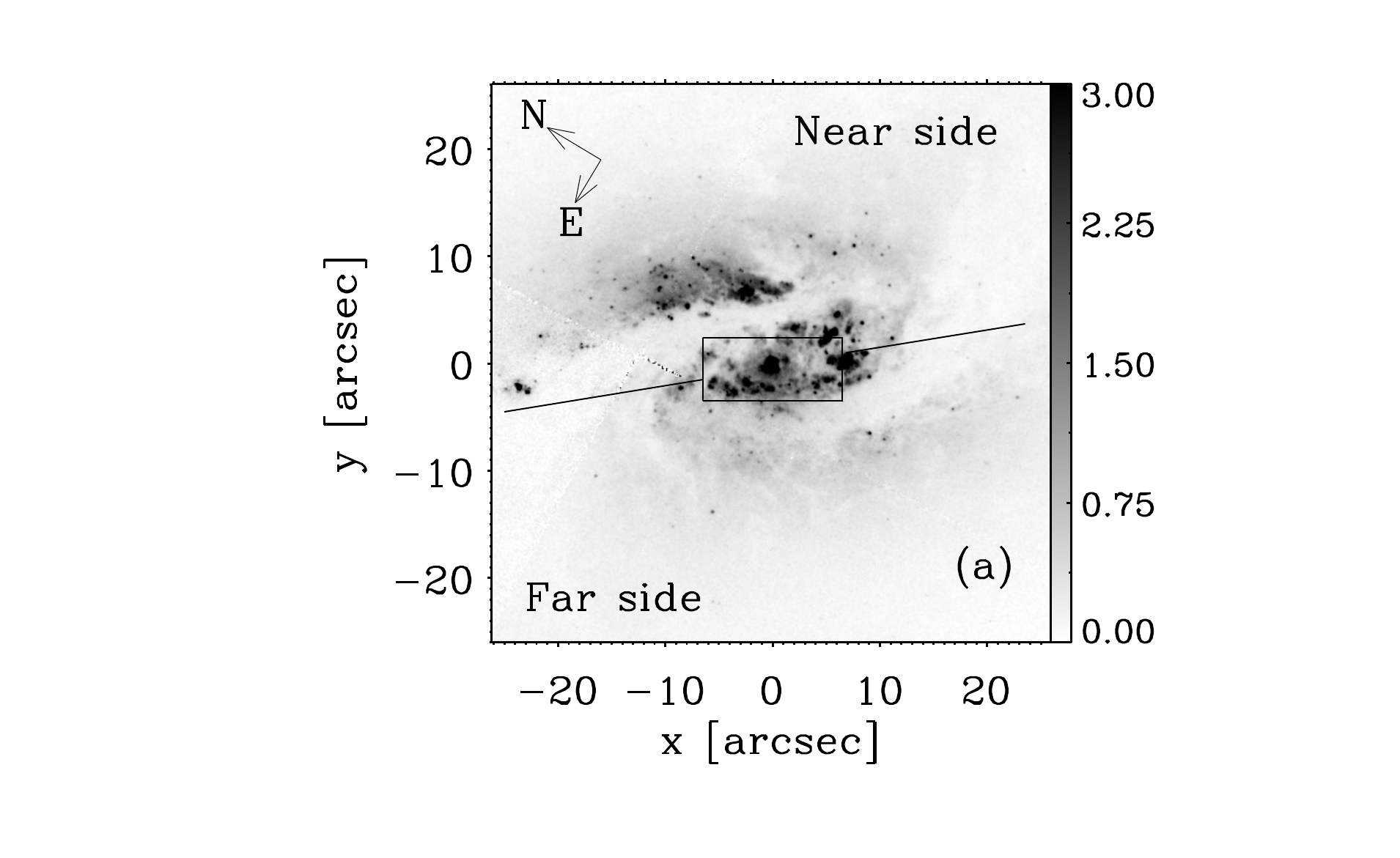}
 \end{minipage}
 \begin{minipage}[t]{.49\linewidth}\vspace{0pt}$
 \begin{array}{c}
   \includegraphics[trim=3.8cm 4cm 4.82cm 2.5cm, clip=true, scale=0.74]{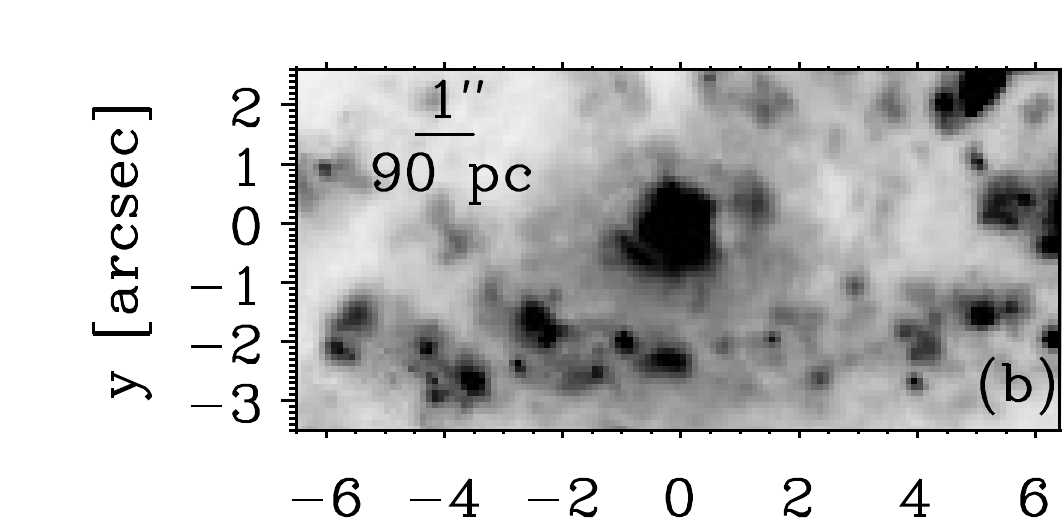}\\
      \includegraphics[trim=-1.1cm 0.5cm 1.9cm 1cm, clip=true, scale=0.42]{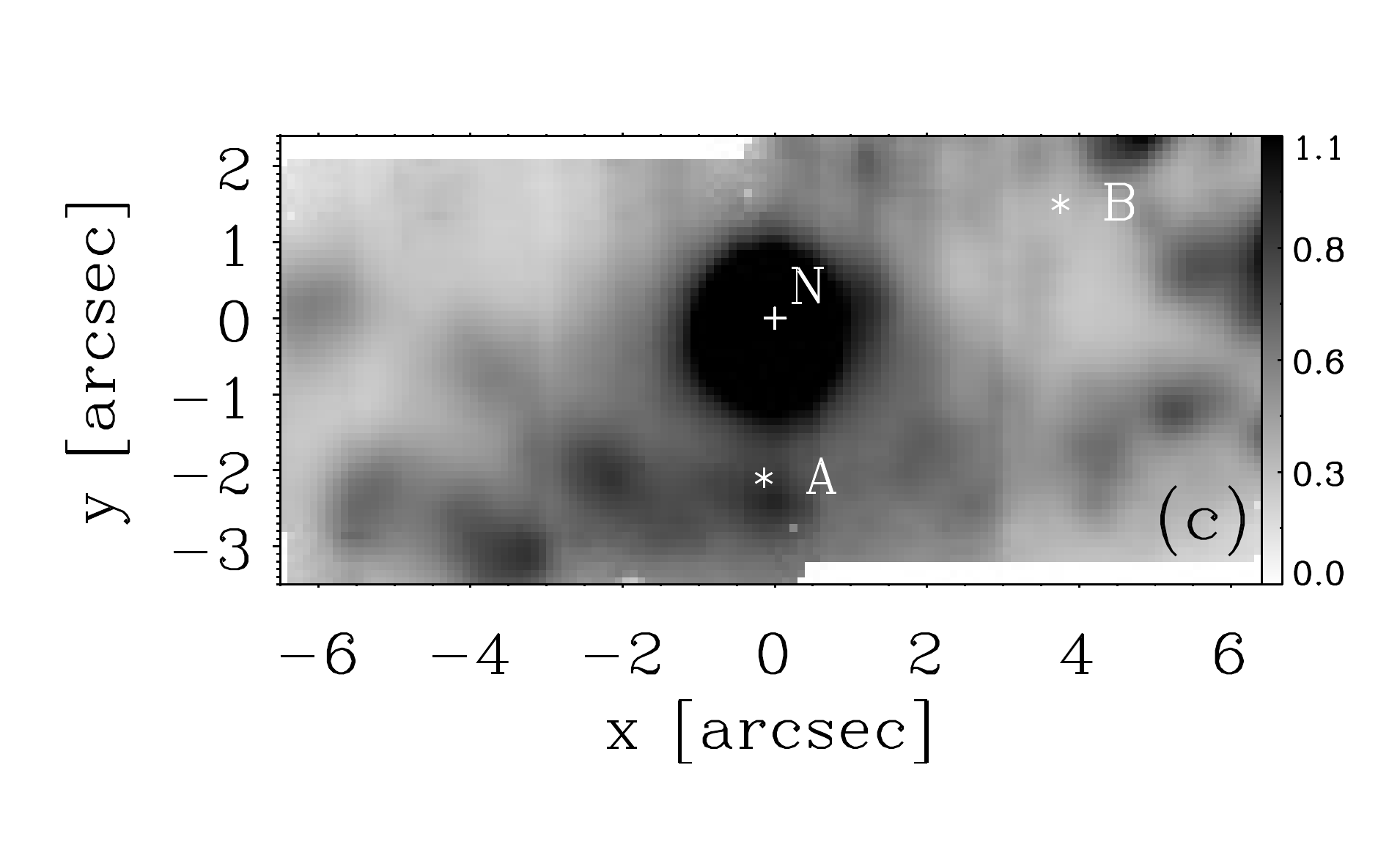}
   \end{array}$
  \end{minipage}
        \subfloat{\includegraphics[trim=0cm 0.25cm 0cm 0.5cm, clip=true, width=1\textwidth]{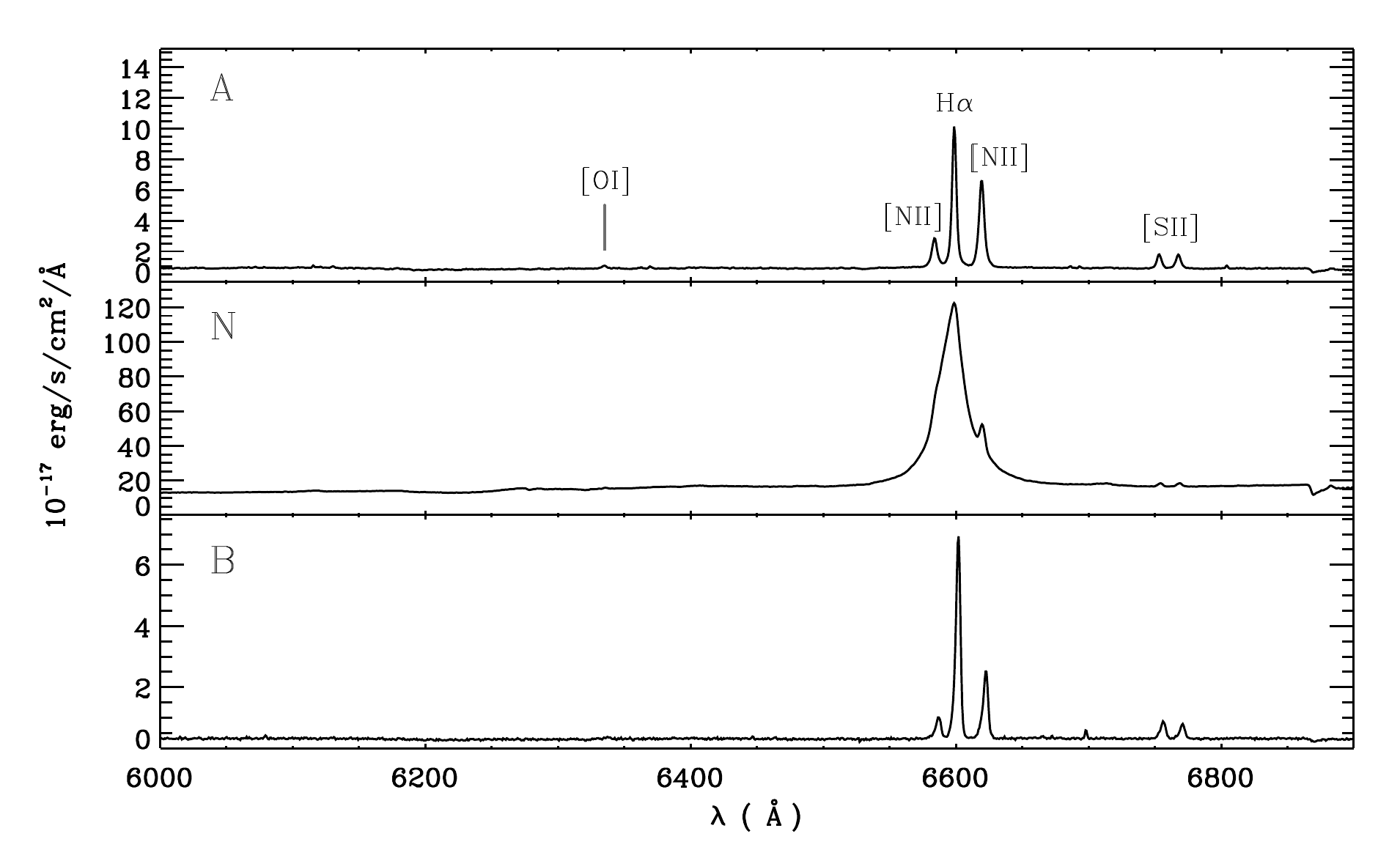}}
  \caption{\textit{Top:} (a) HST image of the nuclear region of NGC 1365 (WFPC2/WF - F547M, PID: 5222). The GEMINI field of view is shown as a box. The solid line indicates the line of nodes derived from the large-scale kinematics of neutral hydrogen \citep{JorsaterM95}. (b) Zoom-in on the HST image matching the GEMINI IFU field of view. (c) Continuum image extracted from the cube obtained with the GEMINI integral field unit. \textit{Bottom:} representative spectra from the spaxels A, B and N, as indicated in the continuum image (c).}
\label{fig:fov}
\end{figure*}

\section{Observations and data reduction} \label{sec: analysis}

NGC 1365 was observed on November 3, 2012 and January 18, 2013 with the Integral Field Unit (IFU) of the Gemini Multi-Object Spectrograph \citep[GMOS;][]{AllingtonSEtAl2002, HookEtAl2004} mounted on the Gemini South Telescope (program ID: GS-2012B-Q73). Two adjacent fields were observed, each covering 7 $\times$ 5 arcsec$^{2}$, resulting in a total angular coverage of 13 $\times$ 6 arcsec$^{2}$ with 0\farcs2 sampling, centered on the nucleus, and roughly aligned with the nuclear bar. The observed field-of-view (FOV) is indicated in the top left panel of Fig.\ref{fig:fov}; continuum images are shown in the right panels.

In order to account for dead fibers in the IFU, $\pm$ 0\farcs35 spatial dithering was applied along both axes. This resulted in four exposures of 2700 seconds each.
From two images of the nuclear point source we determined that the spatial resolution is 0\farcs58 $\pm$ 0\farcs1 or 52 $\pm$ 9 pc, where the uncertainty represents a fiducial value for the variation of the seeing during the observation. The images of the nuclear point source were obtained by averaging the cube in the spectral direction within the range $6531$ \AA $< \lambda < 6568$ \AA, and $6632$ \AA $< \lambda < 6656$ \AA, that is the spectral region including, respectively, the blue and the red wing of the broad H${\alpha}$ emission.

In order to cover the wavelength range 5600-7000 \AA, which includes the emission lines [OI]$\lambda$6300, H$\alpha$+[NII]$\lambda\lambda$6548,6583, [SII]$\lambda\lambda$6716,6731 and several stellar absorption features, we used the IFU in two-slits mode with the grating GMOS R400 in combination with the r(650 nm) filter, yielding a spectral resolution R = 1918. From the width of the emission lines in the spectrum of a calibration lamp, we estimate that the actual spectral resolution is $\sigma \approx 1.11$ \AA, which corresponds to $\sigma \approx 49$ km s$^{-1}$ within the range of wavelengths where the [NII]$\lambda$6583 emission line is observed.
A portion of the spectrum including the most prominent emission lines is shown in the bottom panel of Fig.\ref{fig:fov} for some representative positions in the FOV.

To perform data reduction we followed the procedure described in detail by \citet{Lena14}. We used the IRAF\footnote{IRAF is the Image Reduction and Analysis Facility, a general purpose software system for the reduction and analysis of astronomical data. IRAF is written and supported by the National Optical Astronomy Observatories (NOAO) in Tucson, Arizona. NOAO is operated by the Association of Universities for Research in Astronomy (AURA), Inc. under cooperative agreement with the National Science Foundation.} packages provided by the GEMINI Observatory, and specifically developed for the GMOS instrument\footnote{\url{http://www.gemini.edu/sciops/data-and-results/processing-software?q=node/11822}}. The process includes bias and sky subtraction, flat-fielding, trimming, wavelength and flux calibration, building of the data cubes, final alignment and combination of the four cubes. 
The final data cube has a spatial binning of $0\farcs1 \times 0\farcs1$ and contains 7611 spectra. The fits to the strongest emission lines were performed with this binning.
However, in order to fit the [OI] line, the signal-to-noise ratio was increased by re-binning to $0\farcs2 \times 0\farcs2$, yielding 1920 spectra\footnote{Of the maps presented in this paper, only the [OI]/H$\alpha$ map (Fig.\ref{fig: ratios}) was derived from the cube binned to $0\farcs2 \times 0\farcs2$.}.
As the observations were carried out over a span of 10 weeks, heliocentric velocity corrections were applied to each cube before combination. 

The typical accuracy in the wavelength calibration is 0.14 \AA\ (or 6 km s$^{-1}$ at $\lambda$ = 6583 \AA). 
Absolute flux calibration is uncertain because observations were obtained over a time span of three months, while only one observation of the standard star was performed. To verify the flux calibration, we compared the H$\alpha$ flux derived here with the results of \citet{SchulzKS94}. In their Table 3, they present the fluxes for both the narrow and broad H$\alpha$ components, as derived with a spatial resolution of $3^{\prime\prime}$, within an aperture of 2\farcs6 $\times$ 6\farcs5, oriented along the east-west direction and approximately centered at 0\farcs5 north of the nucleus. However, \citeauthor{SchulzKS94} stress that in their work there was ``no safely accurate positioning of the slit on the nucleus". With this caveat, we note that they obtained $27$ and $51 \times 10^{-14}$ erg s$^{-1}$ cm$^{-2}$ for the fluxes of the narrow and the broad component respectively. From the flux maps presented here, we obtain fluxes of 33 and 107 $\times 10^{-14}$ erg s$^{-1}$ cm$^{-2}$ for the two components. These quantities were measured within a box which reproduces, approximately, the settings adopted by \citeauthor{SchulzKS94}.

The flux of the narrow component is in good agreement with the result of \citeauthor{SchulzKS94}, while that derived for the broad component is larger by a factor of 2. The difference in the broad H$\alpha$ flux is not too surprising since it can plausibly be attributed to intrinsic variability (AGN broad lines are well known to vary on timescales of weeks to months, e.g. \citealt{Peterson88}); the difference in spatial resolution and the uncertain positioning of the slit in \citeauthor{SchulzKS94}'s observations may also contribute. The flux of the narrow component, on the other hand, should not show any strong variability over time, and it is not strongly affected by the slit position. Therefore, we consider the flux of the narrow component as the most reliable indicator of the flux calibration, and we conclude that the calibration adopted here compares well with the one used by \citeauthor{SchulzKS94}.

\section{Emission line fitting} \label{sec: fitting}
To model the continuum and the profiles of the most prominent emission lines (H$\alpha$, [NII]$\lambda\lambda$6548,6583, [SII]$\lambda\lambda$6716,6731, and [OI]$\lambda$6300), we used a customized version of the IDL\footnote{IDL, or Interactive Data Language, is a programming language used for data analysis and visualization.} routine PROFIT \citep{Riffel10}.
Either Gaussian or Gauss-Hermite profiles were fitted to the lines in order to derive centroid velocities, velocity dispersions and fluxes.
Three separate runs were performed to fit the H$\alpha$ + [NII] lines, the [SII] doublet, and the [OI] line.
The continuum adjacent to the lines was modeled with a first order polynomial. 

Despite the presence of several bright star-forming regions, it is difficult to derive information on the stellar kinematics: over the region where the AGN point-spread-function (PSF) dominates, the AGN continuum is much stronger than the underlying stellar component. Moreover, within the covered spectral range, the only prominent absorption lines due to a young stellar population are the Na I doublet, and H$\alpha$. The first one is contaminated by neutral gas; the second one is hidden under the strong H$\alpha$ emission due to the ionized gas. Because of this, the fit of a stellar template would be poorly constrained and we chose not to do it. In principle, the H$\alpha$ flux should be corrected for the underlying stellar absorption, but, because the H$\alpha$ emission line is very strong, we estimate that this correction is smaller than the error in the line flux (\textsection\ref{subs: flux_err}).

\subsection{Fitting procedure}
To fit the H$\alpha$ + [NII] emission lines we assumed that (i) the narrow component of H$\alpha$ and [NII]$\lambda$6548 have the same width as [NII]$\lambda$6583; (ii) the [NII]$\lambda$6548 emission line has the same redshift as [NII]$\lambda$6583; (iii) the amplitude of [NII]$\lambda$6548 is 1/2.96 times the [NII]$\lambda$6583 amplitude \citep[e.g.][]{OsterbrockF06_book}. A similar approach was used to fit the [SII] doublet: we assumed that both lines have the same redshift and width, however the amplitude of each line was left as a free parameter.

\begin{figure*}$
\begin{array}{cc}
\includegraphics[trim=0cm 0cm 0cm 0.5cm, clip=true, scale=0.65]{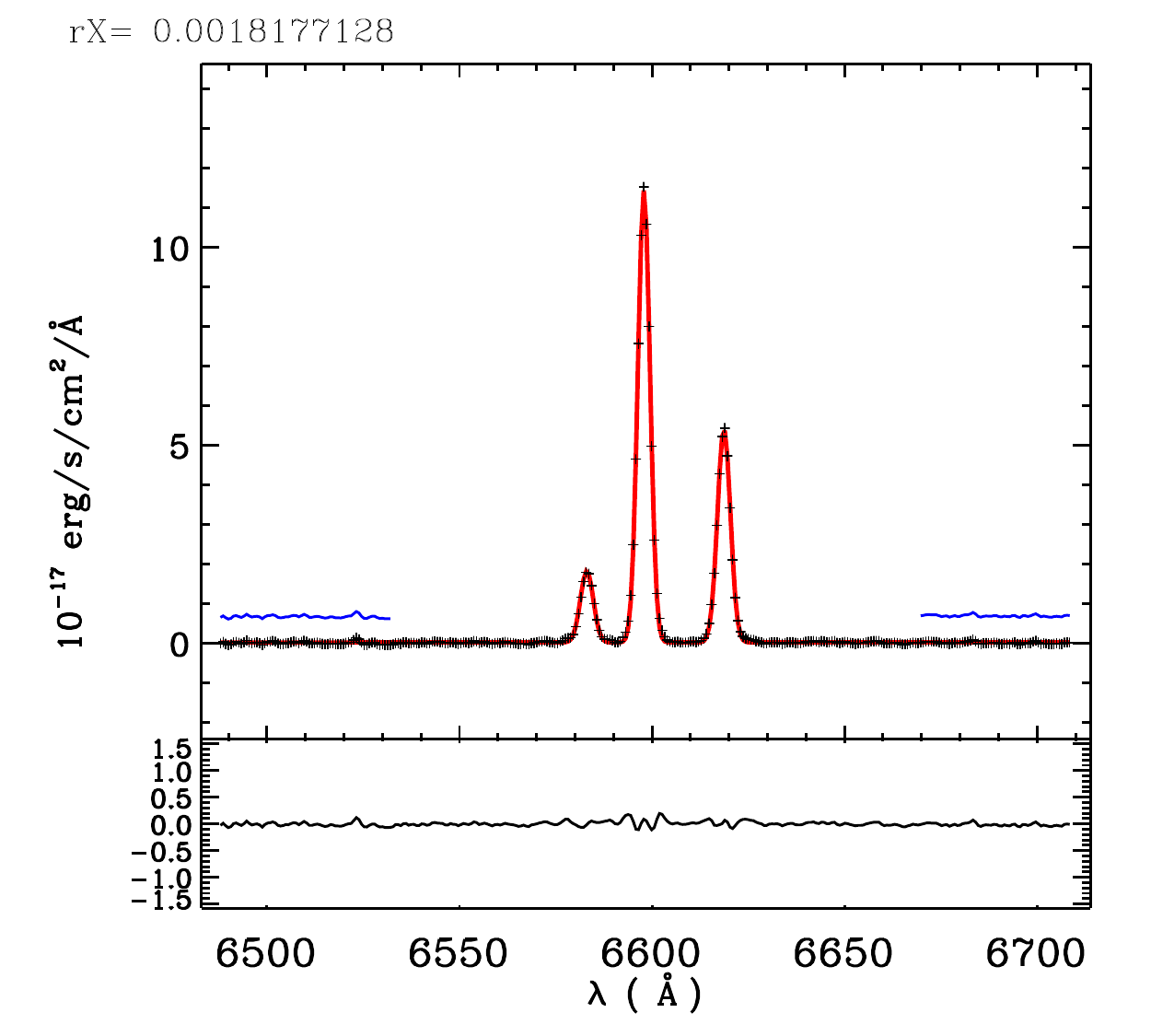} & 
\includegraphics[trim=0.5cm 0cm 0cm 0.5cm, clip=true, scale=0.65]{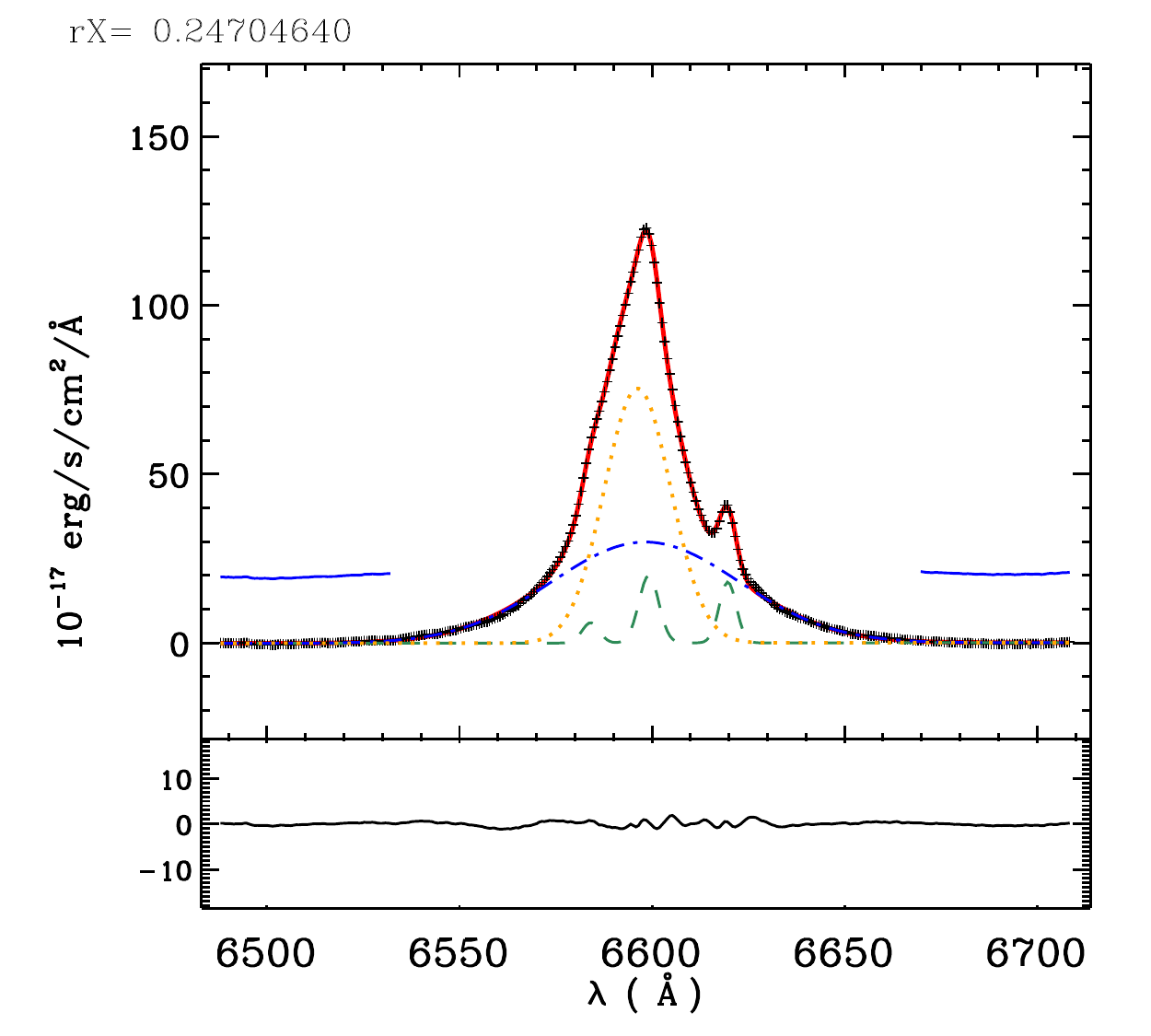}\\
\includegraphics[trim=0cm 0cm 0cm 0.5cm, clip=true, scale=0.65]{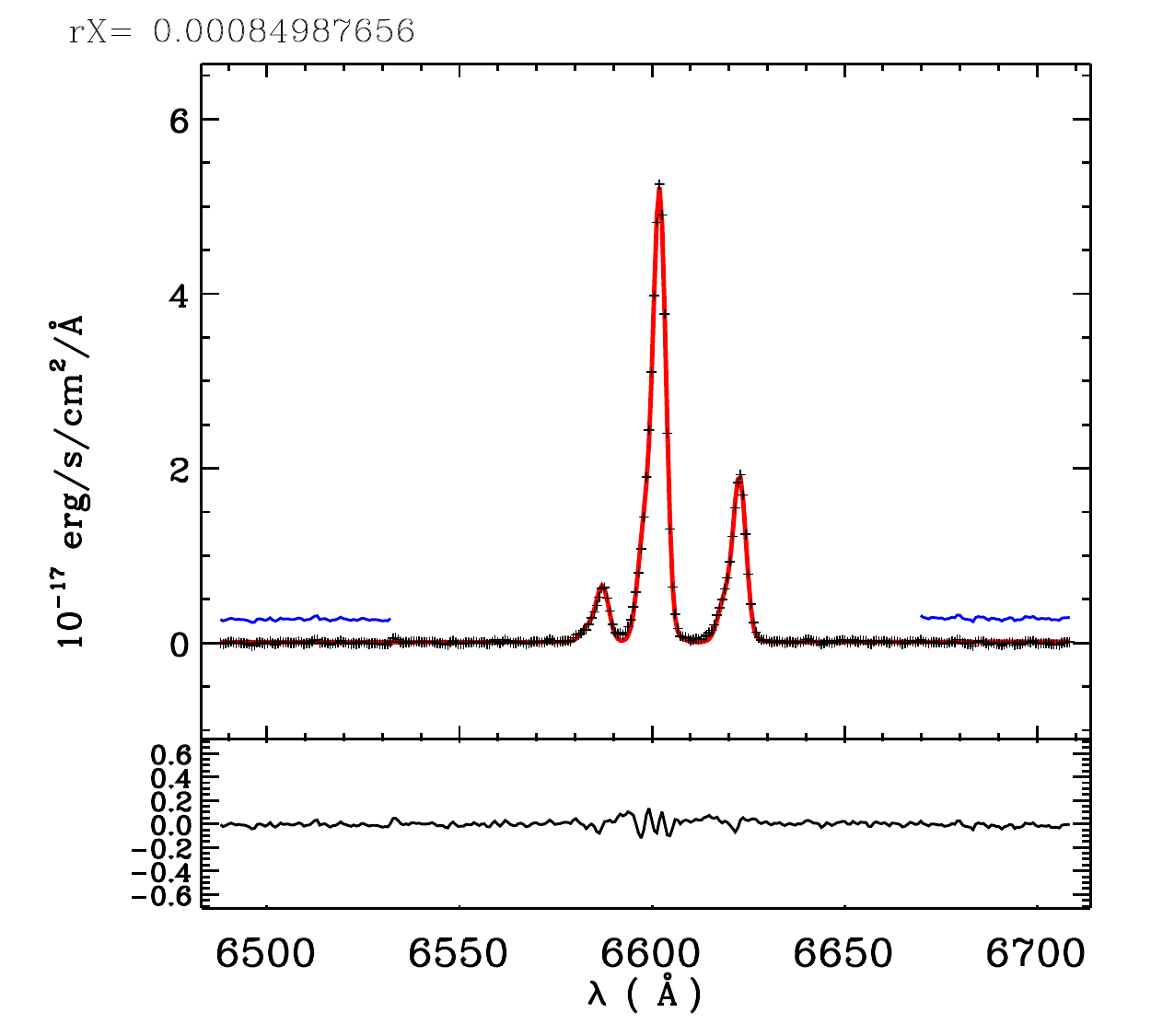} &
\includegraphics[trim=0.5cm 0cm 0cm 0.5cm, clip=true, scale=0.65]{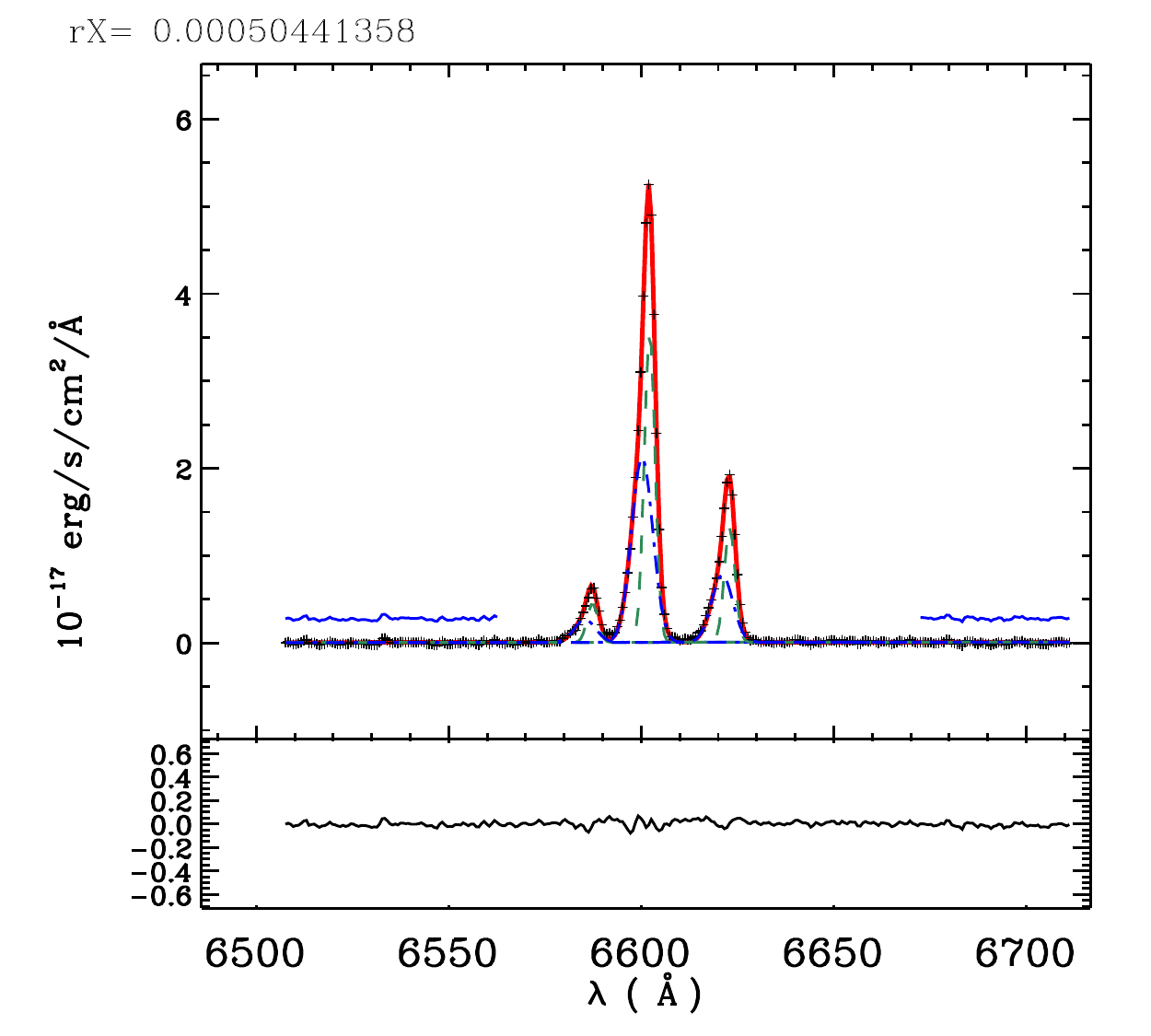}\\
\end{array}$
\caption{Fit of the [NII]$\lambda\lambda$6548,6583 and H$\alpha$ emission lines. \textit{Top left:} example of single Gaussian fit. The horizontal blue lines mark the continuum level before it was subtracted. \textit{Top right:} fit at the nucleus. Two Gaussians were used to model the broad H$\alpha$ (blue dot-dashed and orange short-dashed). The green long-dashed Gaussians represent the underlying narrow components. \textit{Bottom left:} example of a fit with Gauss-Hermite polynomials. \textit{Bottom right:} same spectrum shown in the previous panel, but in this case modelled with two Gaussians to extract the velocity of the component associated with the asymmetry (blue dot-dashed Gaussian).}
\label{fig: fits}
\end{figure*}

Over most of the FOV, the emission lines are well represented by a simple Gaussian profile, e.g. top left panel in Fig.\ref{fig: fits}, however a bright broad H$\alpha$ line is present at the nucleus, e.g. spectrum N in the bottom panel of Fig.\ref{fig:fov}, and the emission lines exhibit asymmetric bases over certain regions of the FOV, e.g. spectrum B in the bottom panel of Fig.\ref{fig:fov}. The strategies adopted to fit the emission line profiles are described below for each case.
\vskip5pt

\textit{The broad line region (BLR):} the central $0\farcs6$ is dominated by unresolved emission from the BLR which manifests itself in the form of a broad H$\alpha$ emission line. At the nucleus it dominates the underlying narrow component and is still visible as a broad base up to $2^{\prime\prime}$ from the nucleus, tracing the extended wings of the PSF.

As the BLR is unresolved, we fitted it with a combination of two Gaussians for which only the total flux was left as a free parameter. We used three spaxels located at different positions around the nucleus at a radius of approximately 0\farcs5 to determine the width, velocity and relative flux of the two Gaussians. In these spaxels the narrow emission lines associated with [NII] and H$\alpha$ are strong enough with respect to the broad component to robustly constrain the fit to the line profile. The mean values of the parameters derived from these fits are shown in Table \ref{tab: blr}. A fit performed at the nucleus using these parameters is shown in the top right panel of Fig.\ref{fig: fits}. The underlying narrow emission lines corresponding to [NII] and H$\alpha$ were fitted simultaneously. 

\textit{Asymmetries (Gauss-Hermite polynomial fit):} asymmetric line profiles are present in some regions of the FOV. To map the spatial distribution of the asymmetry we fitted the emission lines with a truncated Gauss-Hermite polynomial (GHP):
\begin{align}
f_{gh}(\lambda) & = F\frac{e^{-k^{2}/2}}{\sqrt{2\sigma^{2}}} \left(1 + h_{3}H_{3} + h_{4}H_{4}\right)\\
H_{3} & = \frac{1}{\sqrt{\sigma}} \left(2\sqrt{2}k^{3} - 3\sqrt{2}k\right) \\
H_{4} & = \frac{1}{\sqrt{24}} \left(4k^{4} - 12k^{2} + 3\right)\\
k & = \frac{\left(\lambda - \bar \lambda\right)}{\sigma},
\end{align}
where $F$ is the flux, $\sigma$ the velocity dispersion, and $\bar \lambda$ is the peak wavelength. Departures from symmetry are quantified by the value of the coefficient $h_{3}$, a proxy for the skewness: a negative (positive) value of this coefficient indicates the presence of a blueshifted (redshifted) tail. The coefficient $h_{4}$, a proxy for the kurtosis, measures the degree of ``peakiness" of the emission line: a negative (positive) value of this coefficient indicates a boxy (centrally-peaked) line profile. An example of a spectrum fitted with a GHP is shown in the bottom left panel of Fig.\ref{fig: fits}.

It is reasonable to argue that asymmetries in the line profile originate from a superposition of multiple kinematical components (e.g. gas rotating in a disk plus gas experiencing in/outflow). While the GHP allows us to derive accurate fluxes for the total line profile, it does not disentangle the multiple components. To achieve this goal, we also fit the asymmetric line profiles with two Gaussians.

\textit{Asymmetries (two Gaussians fit):} to derive flux, velocity dispersion and velocity for the kinematical component responsible for the observed asymmetry, we fitted two Gaussians to those lines where the skewness coefficient satisfies the condition $|h_{3}| \geq 0.05$. The condition was determined by trial and error to select those regions where the asymmetry is strong enough to allow a robust fit with two Gaussians. An example of a two-components fit is shown in the bottom right panel of Fig.\ref{fig: fits}.

\subsection{Uncertainties}
\label{subsec: uncert_N1365}
To estimate the errors on the measured quantities we performed Monte Carlo simulations: for each spaxel, we constructed one hundred realizations of the spectrum by adding Gaussian noise with amplitude comparable to the noise measured in the original spectrum. Mean values and standard deviations for the centroid velocities, velocity dispersions and fluxes were derived for each spaxel, with the standard deviation of the distribution in each parameter being taken as the uncertainty.


\begin{table}
\caption{THE BROAD COMPONENTS}  \label{tab: blr}
\scalebox{0.85}{
\begin{tabular}{c lll}
\hline
\hline
 & &   \\
component		& vel [km s$^{-1}$]			& $\sigma$ [km s$^{-1}$] 	& relative flux (amplitude)	\\
	 			&  						& 						\\
\hline
 & &   \\
	1 			&  -143 $\pm$ 1			& 394 $\pm$ 1				& 1 (1)\\
	2 			&  -51 $\pm$ 2 				& 1113 $\pm$ 2			& 0.893 (0.411)\\
	tot			& -100 $\pm$ 2$^{*}$		& 1181 $\pm$ 2$^{\dagger}$	& $\ldots$\\
 & & \\
\hline
\end{tabular}}
Notes -- $^{*}$ This is a flux weighted average of the velocities derived for the two components. All velocities have been computed after subtracting a systemic velocity of 1671 km s$^{-1}$. $^{\dagger}$ This value is computed as $\sqrt{\sigma_{1}^{2} + \sigma_{2}^{2}}$. Fluxes are free parameters with a different value for each pixel. As a reference, for both Gaussians the median flux per pixel within a radius of 0\farcs3 from the continuum peak is 1.2 $\times$ 10$^{-14}$ erg s$^{-1}$ cm$^{2}$, where 1 pixel $\equiv$ 0\farcs1 $\times$ 0\farcs1.
\end{table}

\section{Results} \label{sec: results}
\subsection{Velocity and velocity dispersion}
\label{subsec: vel_N1365}
\begin{figure}
\includegraphics[trim=0.7cm 1.8cm 0cm 0.5cm, clip=true, scale=0.51]{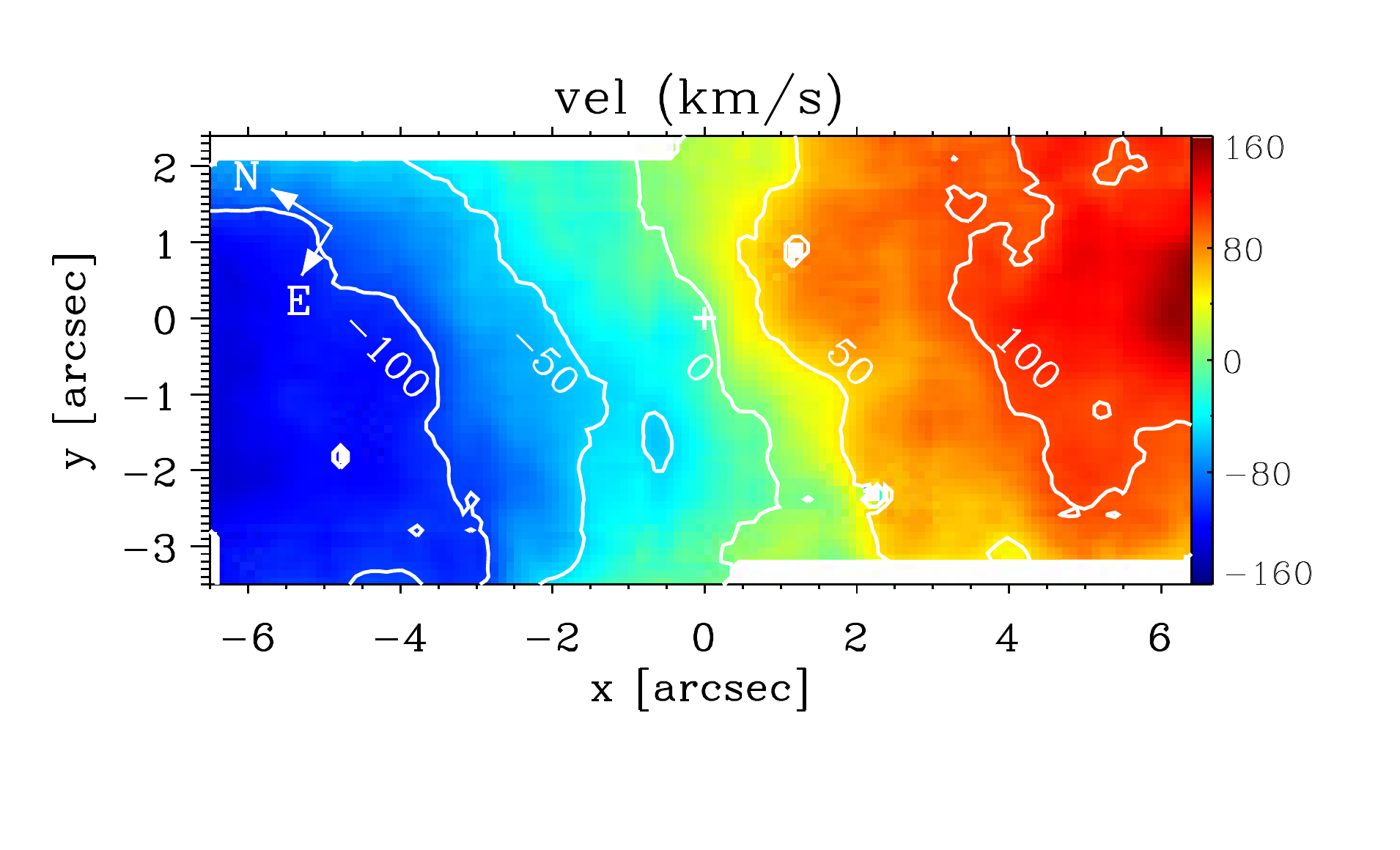}\\
\includegraphics[trim=0.7cm 1.8cm 0cm 0.5cm, clip=true, scale=0.51]{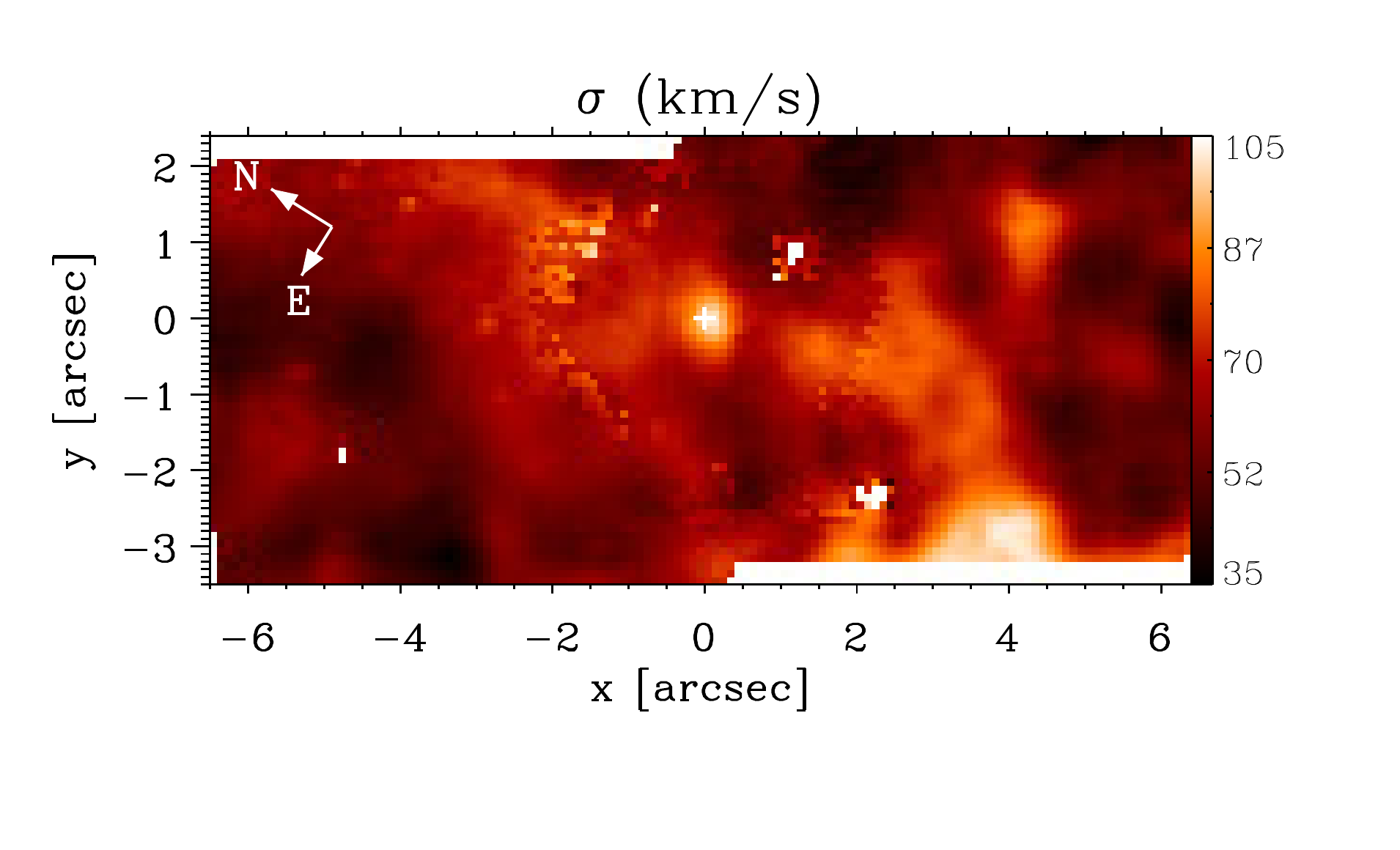}\\
\includegraphics[trim=1.7cm 1.8cm 0cm 0.5cm, clip=true, scale=0.51]{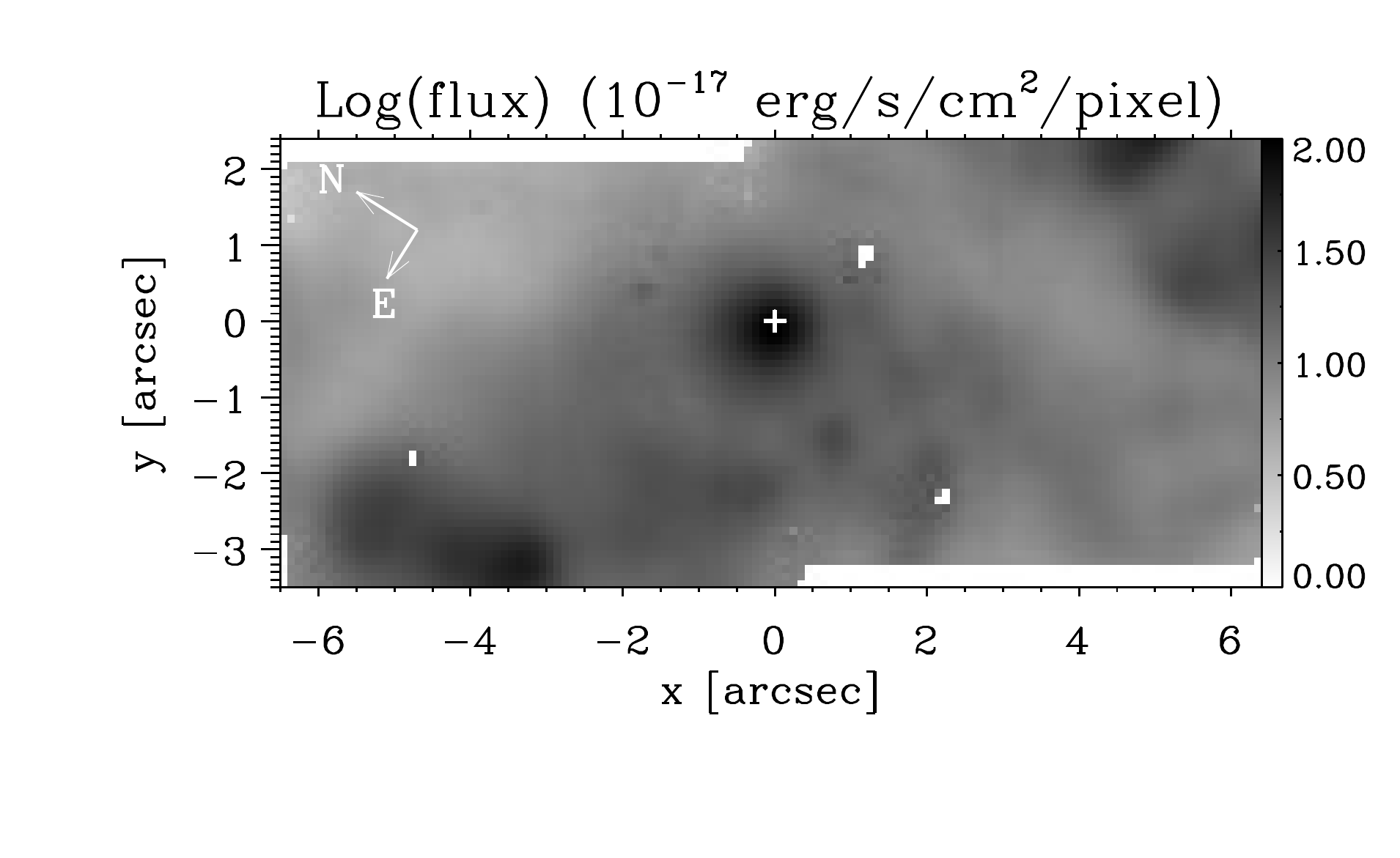}
\caption{Maps derived by fitting a Gaussian profile to the narrow component of the [NII] emission line. \textit{Top}: velocity. \textit{Center}: velocity dispersion. \textit{Bottom}: flux. The velocity map is shown after subtracting a systemic velocity of 1671 km s$^{-1}$ (derived from the modeling described in \textsection \ref{subsec: gasv_model}). The cross at (0,0) marks the continuum brightness peak.}
\label{fig: vel_narrow}
\end{figure}

The centroid velocity and velocity dispersion derived from the narrow [NII]$\lambda$6583 emission line are shown in Fig.\ref{fig: vel_narrow}. The maps were produced by combining results from the single-Gaussian fit and the multiple component fit to the spectra where a broad H$\alpha$ line is present. 

The intrinsic velocity dispersion was derived as:

\begin{align} 
\sigma_{int} = \sqrt{\sigma_{obs}^{2} - \sigma_{ins}^{2}},
\end{align}

\noindent where $\sigma_{obs}$ is the observed velocity dispersion, as derived from the fit of the emission lines, and $\sigma_{ins}$ is the instrumental broadening, as derived from the fit of the emission lines in the spectra of a calibration lamp. Within the range of wavelengths where the [NII]$\lambda$6583 emission line is observed, that is $6610<\lambda<6630$ \AA, $\sigma_{ins}$ can be approximated to $49$ km s$^{-1}$. After the correction is applied, the velocity dispersion ranges between 35 and 105 km s$^{-1}$ with a mean value of 62 km s$^{-1}$ (approximately coincident with the median).

The velocity field shows a pattern which could originate either from rotation or streaming motions along a bar. 
Hints that an additional kinematical component is superposed on the velocity field derived from the narrow component come from asymmetries in the base of the line profiles. A map of the skewness of the emission lines, the $h_{3}$ coefficient derived from Gauss-Hermite polynomial fits, is presented in the top panel of Fig.\ref{fig: h3}. A comparison of this map with the velocity field in Fig.\ref{fig: vel_narrow} makes it clear that the asymmetry is reversed with respect to the velocity field of the narrow component, i.e. the asymmetry is blueshifted (redshifted) where the emission lines are redshifted (blueshifted) with respect to the systemic velocity. In regions where $|h_{3}| \geq 0.05$ each line was modelled with two Gaussians. The centroid velocity and velocity dispersion for the component associated with the asymmetry are presented in Fig.\ref{fig: asym}. 

The BLR is assumed to be unresolved, therefore the kinematical parameters used to fit the broad H$\alpha$ components were not allowed to vary over the FOV. Values for the velocity and velocity dispersion adopted for each Gaussian component are listed in Table \ref{tab: blr}. 

Centroid velocities and velocity dispersions derived from the fits to the H$\alpha$, [SII], and [OI] emission lines are consistent with the values derived from [NII].


\subsubsection{Uncertainties}
Errors due to the fitting procedure, on the velocity and velocity dispersion, were estimated with a Monte-Carlo simulation, as explained in \textsection \ref{subsec: uncert_N1365}.
The uncertainty values derived for the single Gaussian fit are typically lower than 1 km s$^{-1}$ for the centroid velocity of the [NII] and H$\alpha$ narrow lines. Errors are slightly larger for the [SII] lines, but still below 5 km s$^{-1}$, with a median value approximately equal to 1 km s$^{-1}$. Typical errors of 5 km s$^{-1}$ are associated with the velocity derived from the [OI] line. Similar values were obtained for the velocity dispersion.

Typical errors for the asymmetric component are approximately 1 km s$^{-1}$ for both the velocity and velocity dispersion in the right side of the FOV, where the asymmetry is stronger. In the left side, where the asymmetry is weaker, typical errors are approximately 10 km s$^{-1}$.

Errors for the velocity and velocity dispersion of the broad Gaussians used to model the broad H$\alpha$ are given in Table \ref{tab: blr}.

\subsection{Line fluxes}

The flux maps for the [NII], H$\alpha$, [SII], and [OI] emission lines all show roughly the same general structures, although there are differences in detail. For example, the integrated flux distribution for the [NII]$\lambda6583$ emission line is shown in the bottom panel of Fig.\ref{fig: vel_narrow}: the strongest emission is observed at the nucleus; two bright blobs, probably HII regions, are visible approximately 5$^{\prime\prime}$ north-east and south-west of the nucleus. There are also several fainter blobs, like those in a ring-like structure around the nucleus, at a distance of $1^{\prime\prime} - 2^{\prime\prime}$. As the main features are similar, the flux maps for H$\alpha$, [SII], and [OI] are not shown, but these were used to construct the line ratio maps, [NII]/H$\alpha$, [SII]/H$\alpha$ and [OI]/H$\alpha$, which are presented in Fig.\ref{fig: ratios} and discussed in Section \ref{subsec: ratio}.

The flux map for the asymmetric component fitted to [NII] and H$\alpha$ is shown in the bottom panel of Fig.\ref{fig: asym}.

\subsubsection{Uncertainties}
\label{subs: flux_err}

Uncertainties on the fluxes, as derived from the fitting procedure, have been estimated with the Monte Carlo simulation described in \textsection \ref{subsec: uncert_N1365}. Typical uncertainties are below $10\%$ on both fluxes and flux ratios for the strongest lines, i.e. [NII] and H$\alpha$. Typical uncertainties on the fluxes of the weaker lines are approximately 10$\%$, corresponding to an uncertainty of about 15$\%$ on the flux ratio.

\begin{figure}
\includegraphics[trim=0.5cm 1.8cm 0cm 0.5cm, clip=true, scale=0.475]{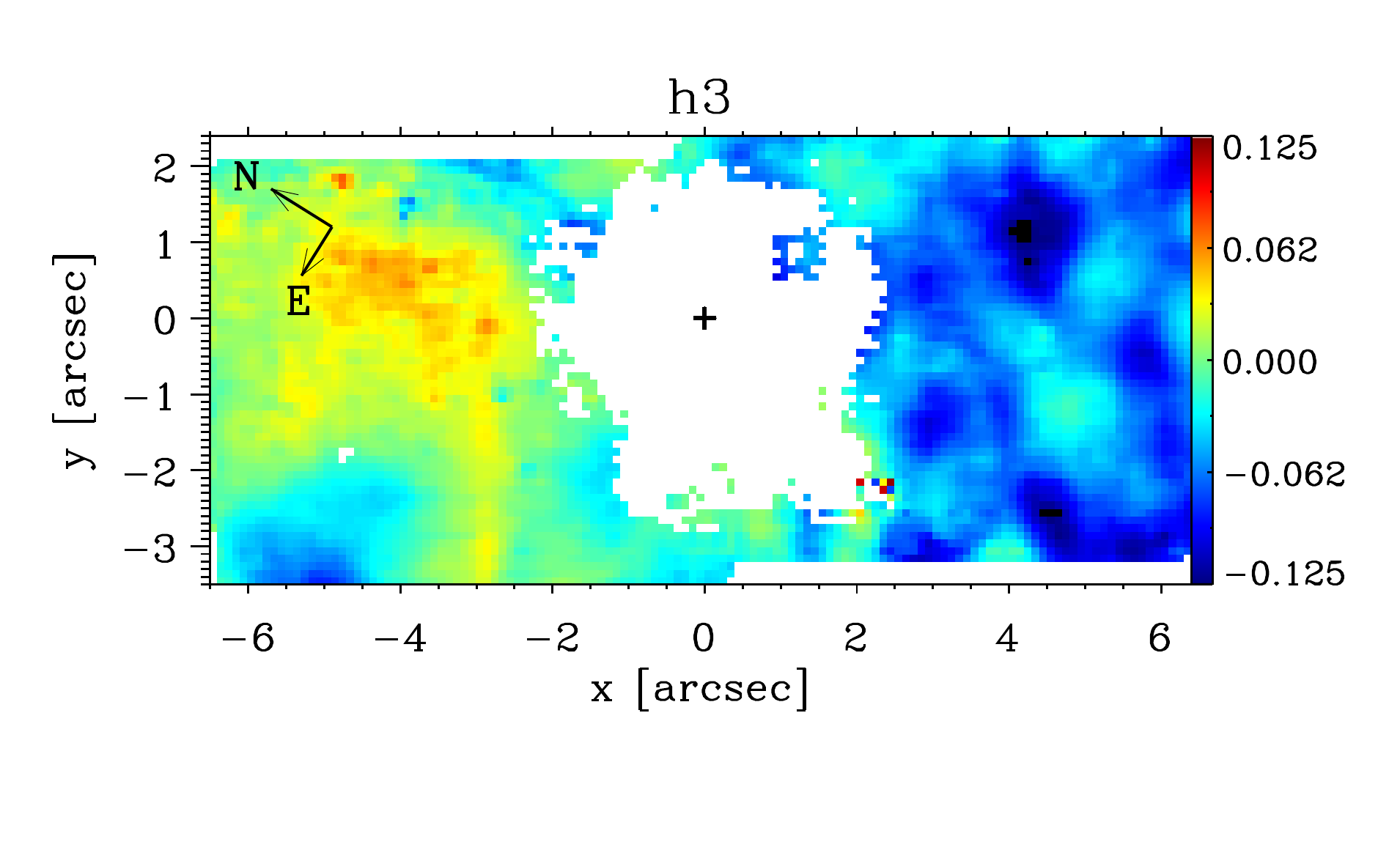}\\
\includegraphics[trim=0.5cm 1.8cm 0cm 0.5cm, clip=true, scale=0.475]{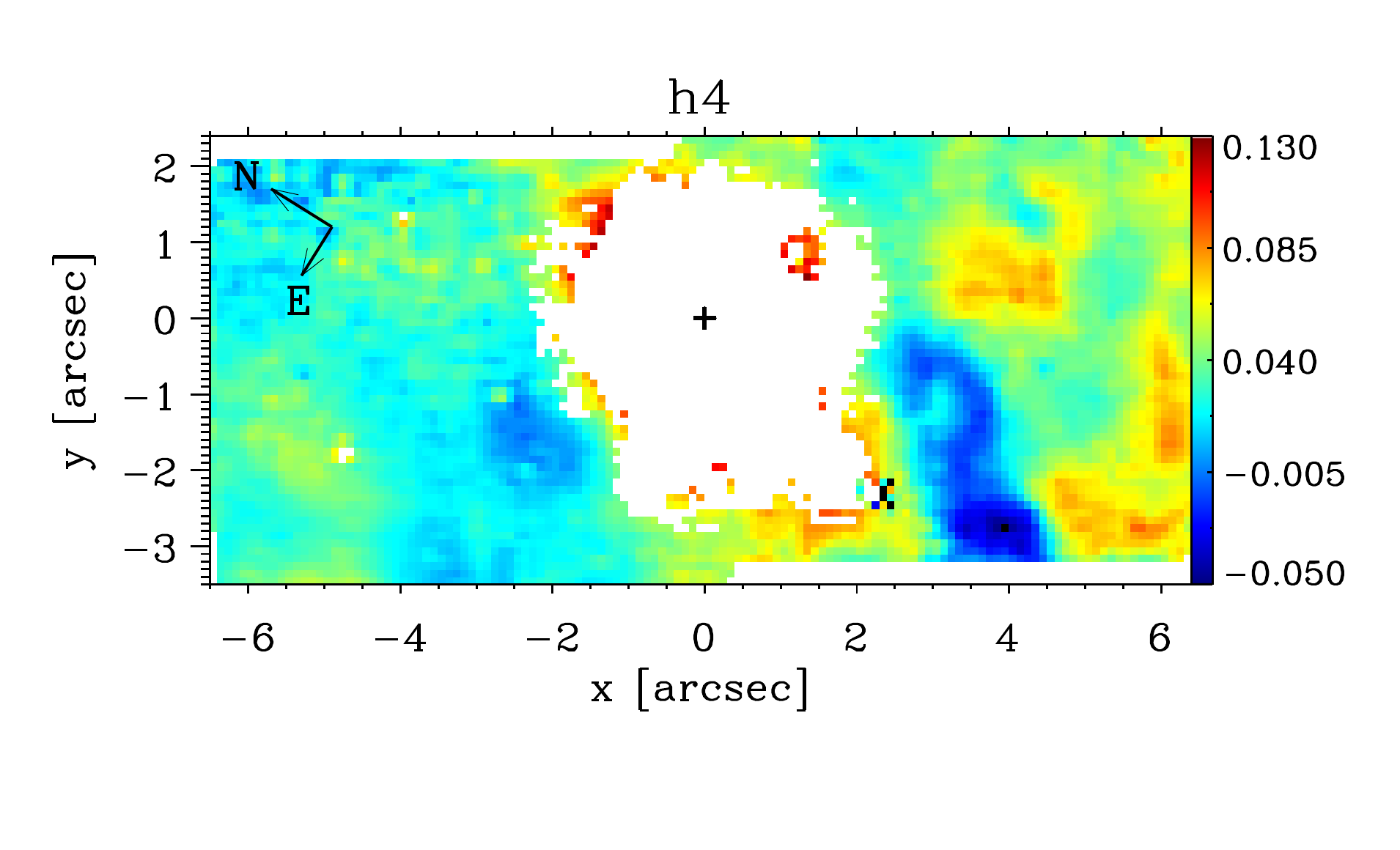}\\
\caption{Maps of the Gauss-Hermite coefficients $h_{3}$ and $h_{4}$. The central region was excluded because of the presence of a broad H$\alpha$ component which did not allow a reliable fit with Gauss-Hermite polynomials. As a reference, a spectrum extracted from the region close to (4\farcs5,1$^{\prime\prime}$), where the asymmetry reaches the maximum, is presented in the bottom panel of Fig.\ref{fig: fits}.}
\label{fig: h3}
\end{figure}

\begin{figure}
\includegraphics[trim=0.7cm 1.8cm 0cm 0.5cm, clip=true, scale=0.475]{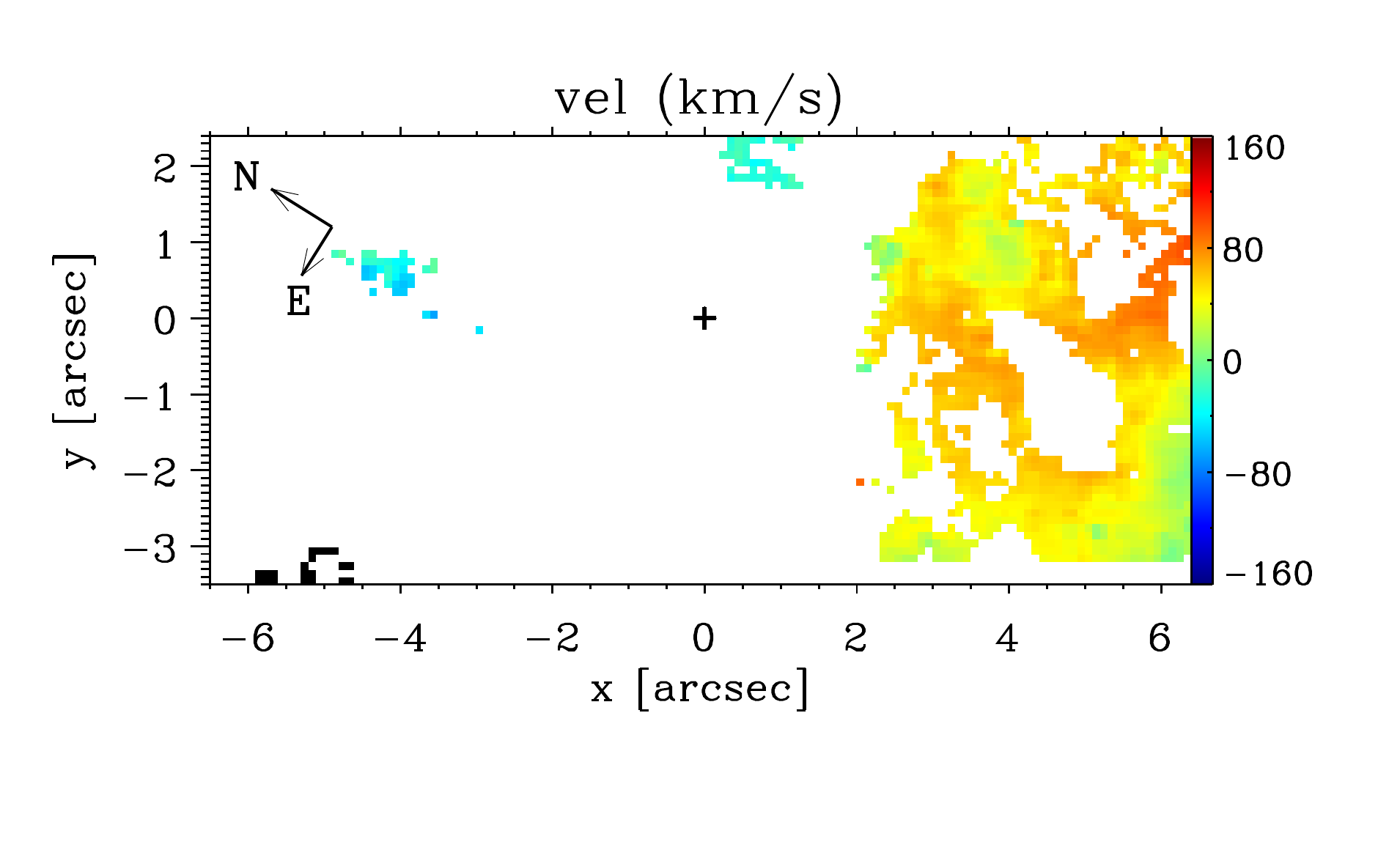}\\
\includegraphics[trim=0.7cm 1.8cm 0cm 0.5cm, clip=true, scale=0.475]{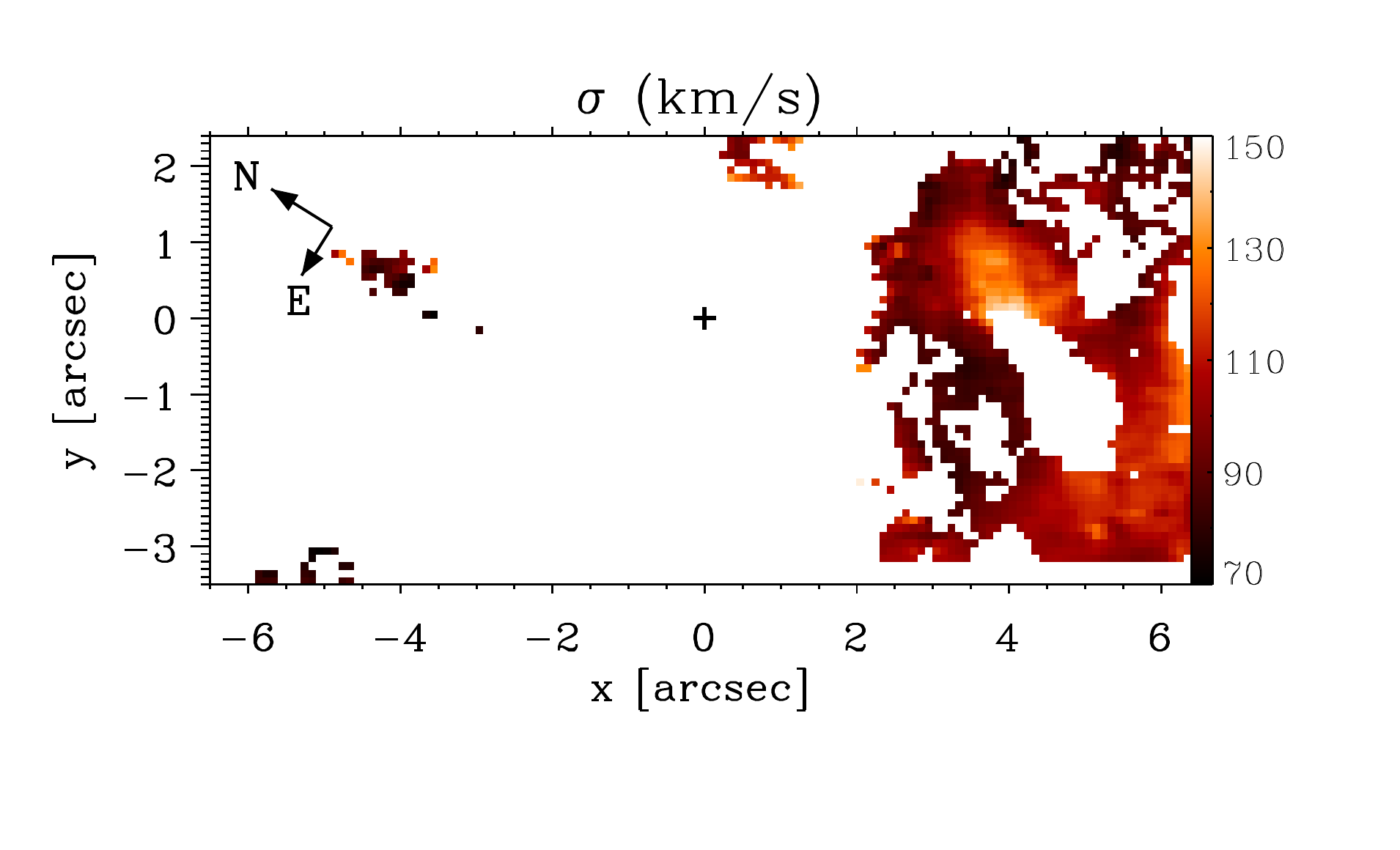}\\
\includegraphics[trim=0.7cm 1.8cm 0cm 0.5cm, clip=true, scale=0.475]{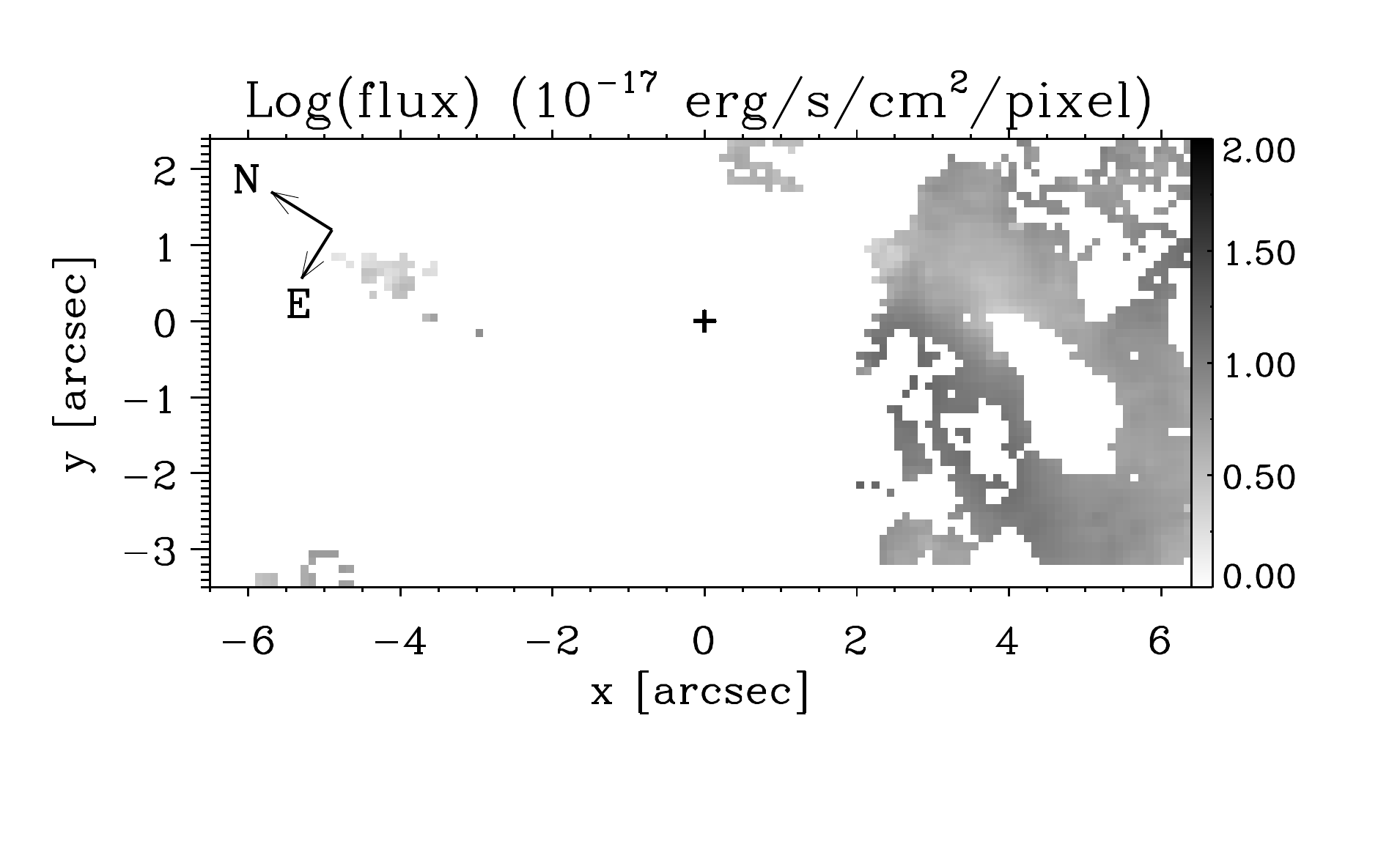}
\caption{Maps derived by fitting two gaussians to the [NII]$\lambda$6583 emission lines characterized by a skewness $|h_{3}| \geq 0.05$. Maps shown here represent the asymmetric component. \textit{Top}: velocity. \textit{Center}: velocity dispersion. \textit{Bottom}: flux. The velocity map is shown ofter subtracting a systemic velocity of 1671 km s$^{-1}$.}
\label{fig: asym}
\end{figure}

\subsection{Electron density}\label{sec: eden}
\begin{figure}
\includegraphics[trim=0.5cm 1.5cm 0cm 0.5cm, clip=true, scale=0.48]{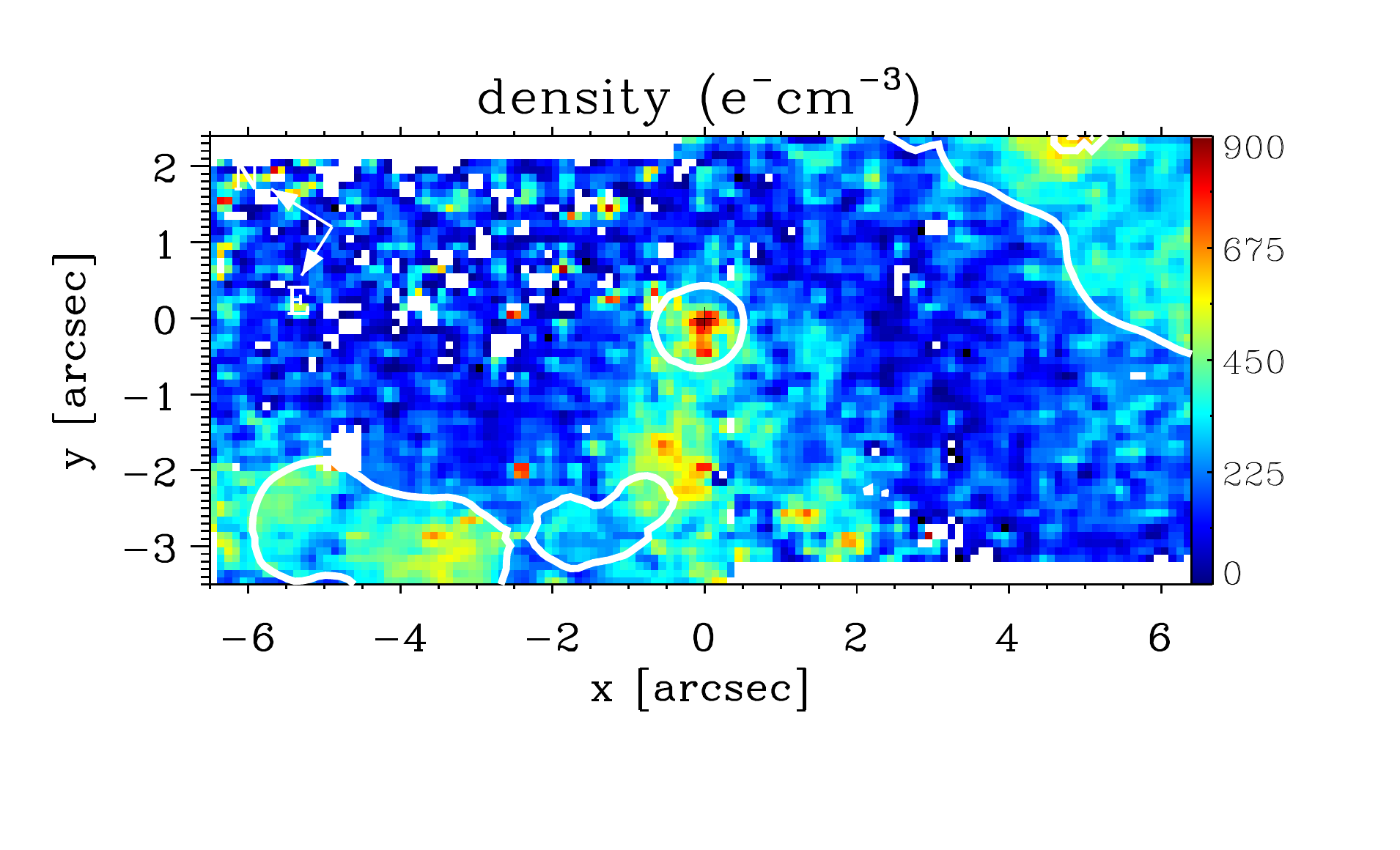}
\caption{Electron density map derived from the flux ratio [SII]$\lambda\lambda$6716/6731 assuming a temperature T = 10$^{4}$ K. The solid white contours mark the most prominent features visible in the H$\alpha$ flux map.}
\label{fig: density}
\end{figure}

The electron density is shown in Fig.\ref{fig: density}. It was derived from the intensity ratio [SII]$\lambda$6716/$\lambda$6731 assuming a temperature of 10$^{4}$K \citep{OsterbrockF06_book}. 

The map shows values in the range 100 - 200 cm$^{-3}$ over most of the FOV, with a typical uncertainty of 60 cm$^{-3}$. The density increases to a mean value of 600 $\pm$ 100 cm$^{-3}$ within a radius of $\approx$ 0\farcs5 from the nucleus. High values ($\approx$ 500 $\pm$ 100 cm$^{-3}$) are present east and south-east of the nucleus, in a region coincident with the cone of [OIII]$\lambda$5007 emission reported by \citet{StorchiBergmannB91}.
Similar high densities are present in the bright blobs, already identified in the flux maps, located about 5$^{\prime\prime}$ north-east and south-west of nucleus.

It was not possible to perform a robust fit of the [SII] doublet with two Gaussians, therefore a density map for the asymmetric component was not derived. 

\subsection{Line ratios}
\label{subsec: ratio}

\begin{figure}
\includegraphics[trim=0.5cm 1.8cm 0cm 0.5cm, clip=true, scale=0.48]{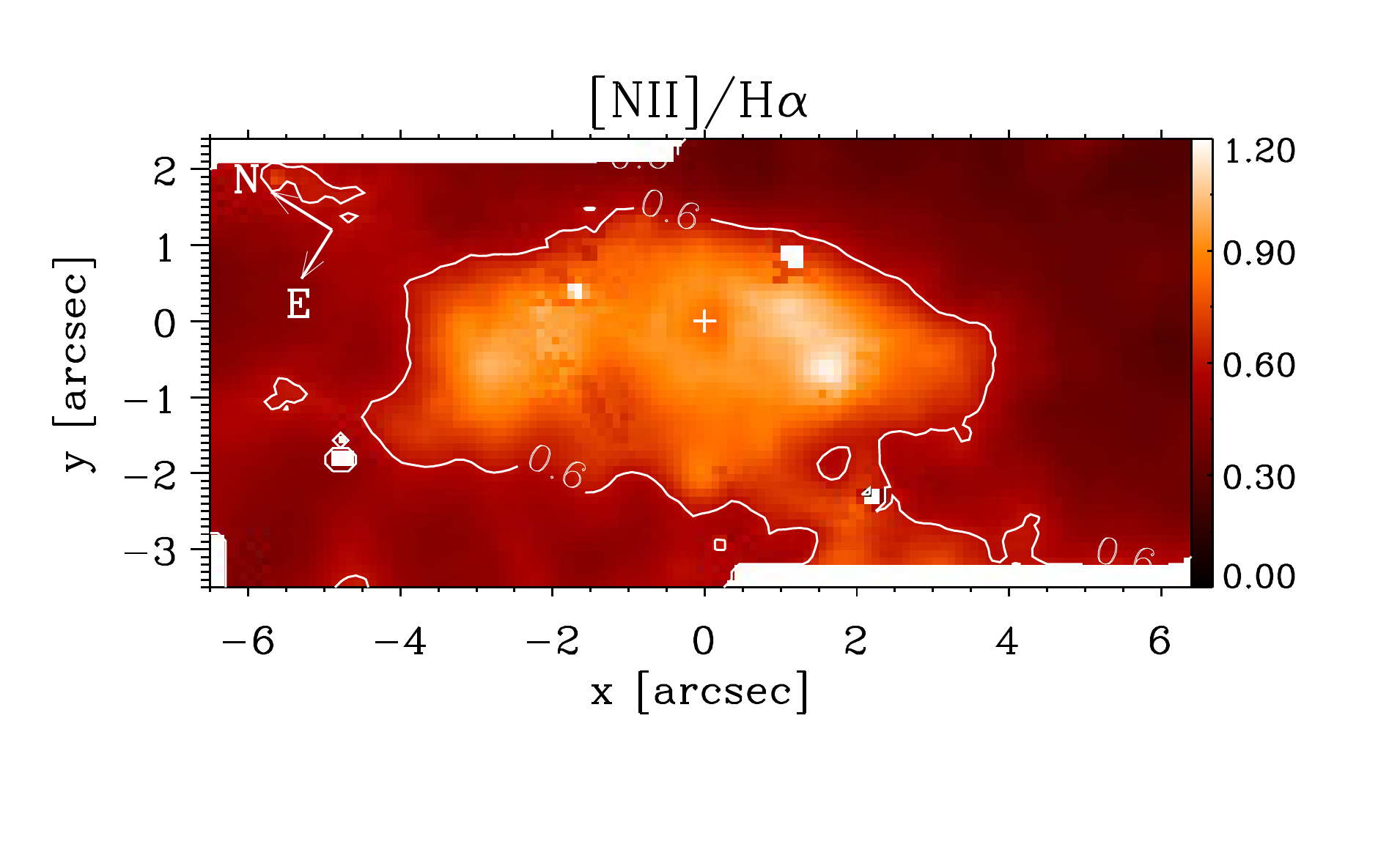}\\
\includegraphics[trim=0.5cm 1.8cm 0cm 0cm, clip=true, scale=0.48]{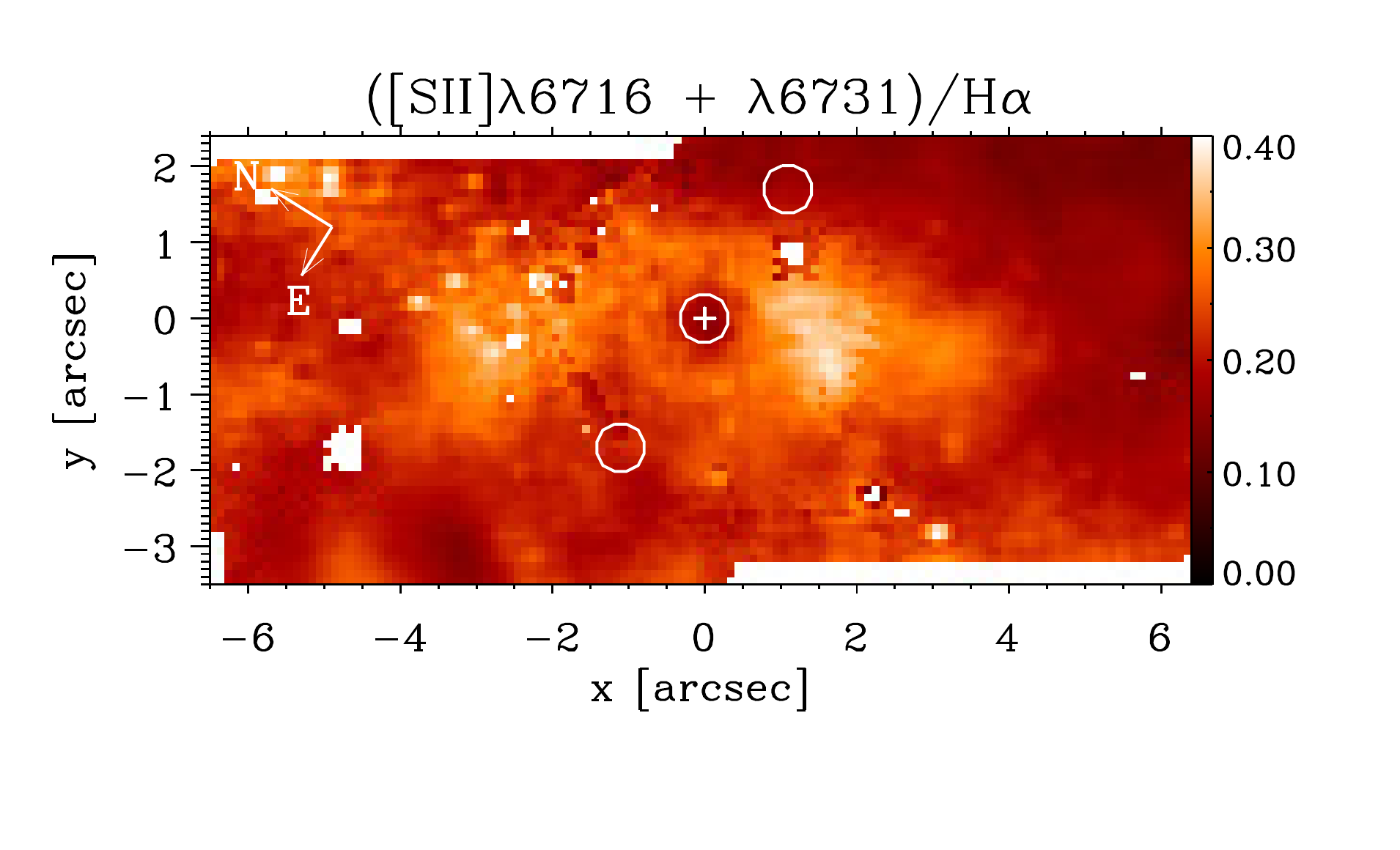}\\
\includegraphics[trim=0.5cm 1.8cm 0cm 0cm, clip=true, scale=0.48]{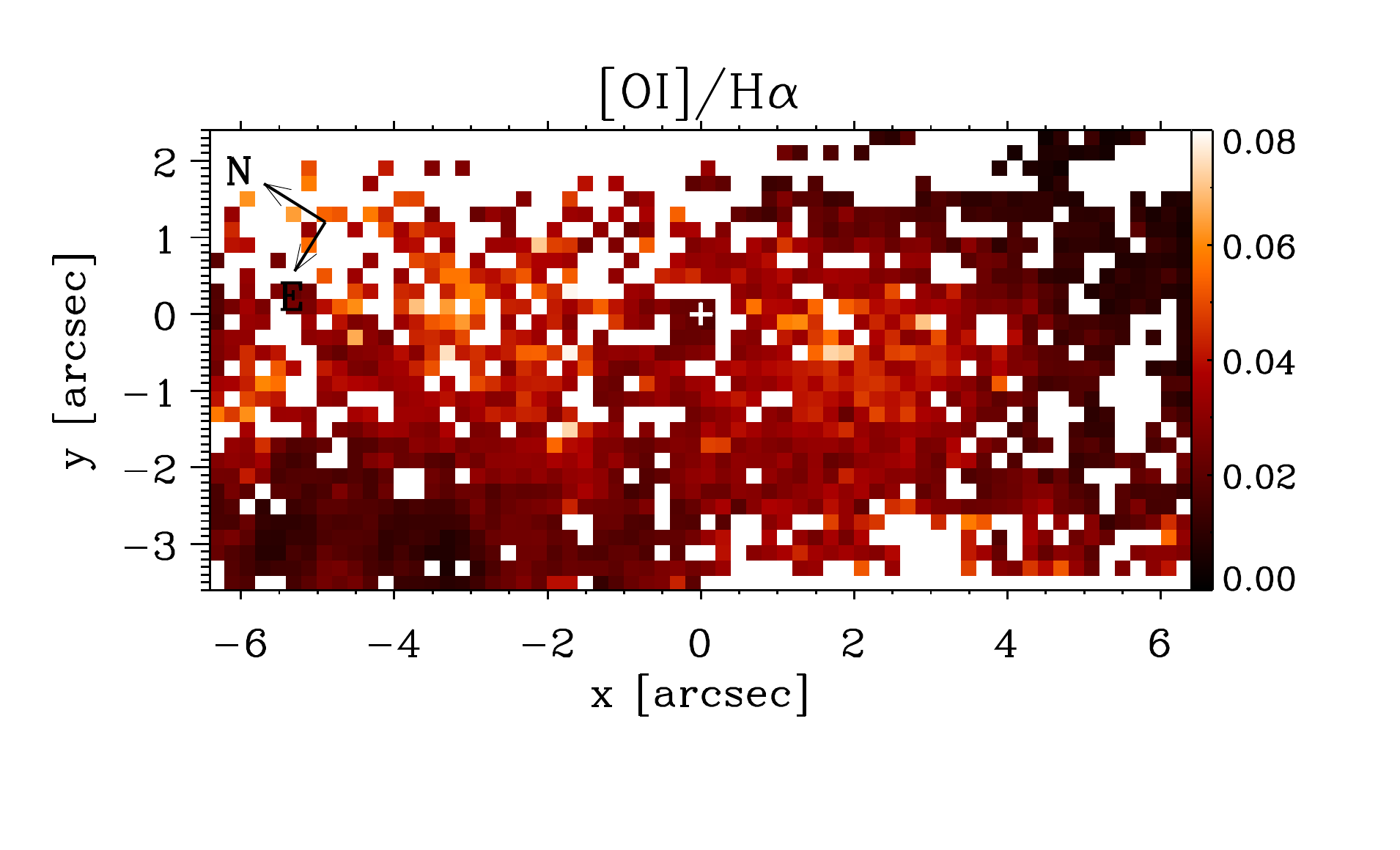}\\
\caption[Flux ratios]{Flux ratios [NII]$\lambda$6583/H$\alpha$, [SII]($\lambda$6716 + $\lambda$6731)/H$\alpha$ and [OI]$\lambda$6300/H$\alpha$. Spaxels where the relative error on the ratio is equal to or larger than 50\% have been masked out (blank pixels). A contour corresponding to [NII]/H$\alpha = 0.6$ is plotted in the top panel; [NII]/H$\alpha > 0.6$ (log 0.6 = -0.2) is characteristic of AGN photoionization. Solid circles in the central panel highlight the regions used to extract the data plotted in the BPT diagrams.}
\label{fig: ratios}
\end{figure}

Maps for the line ratios [NII]$\lambda$6583/H$\alpha$, [SII]($\lambda$6716 + $\lambda$6731)/H$\alpha$ and [OI]$\lambda$6300/H$\alpha$ are presented in Fig.\ref{fig: ratios}. Spaxels where relative errors are equal or larger than 50\% have been masked out.

The maps show high values in an elongated region, roughly centered on the nucleus, extending from -4 to +4$^{\prime\prime}$ along the NE-SW direction, and approximately 1\farcs5 NW and SE of the nucleus, in the orthogonal direction. In this region typical values range between 0.8 (log: -0.1) and 1.2 (log: 0.08) for [NII]/H$\alpha$, between 0.25 (log: -0.6) and 0.36 (log: -0.4) for [SII]/H$\alpha$, and between 0.03 (log: -1.5) and 0.05 (log: -1.3) for [OI]/H$\alpha$. A decrease in the line ratios is visible in the maps within 0\farcs5 from the nucleus, but it is more pronounced in the [SII]/H$\alpha$ ratio, whose values drop into the range 0.15 - 0.24 (log: -0.8, -0.6).
Low values are also visible in a large region approximately west of the nucleus. Values are in the range 0.26 - 0.4 (log: -0.6,-0.4) for the [NII]/H$\alpha$ ratio, 0.12 - 0.17 (log: -0.9,-0.8) for the [SII]/H$\alpha$ ratio, and 0.007 - 0.01 (log: -2.15,-2) for the [OI]/H$\alpha$ ratio.
Intermediate values are present north, east and south of the nucleus.

The increase in the line ratios, in the NE-SW direction, is fairly well defined and evident. It doesn't have any clear corresponding feature in the density or in the kinematical maps, however it is confined within the elongated ring of bright knots clearly visible in the HST continuum images (compare the top panel of Fig.\ref{fig: ratios} with the top right panel in Fig.\ref{fig:fov}). It is possible that the increase in the line ratio traces the region where AGN photoionization dominates over stellar photoionization.

The decrease in the line ratios at the nucleus is larger than the associated uncertainty, especially for the [SII]/H$\alpha$ ratio. However, given the complexity of the spectral profiles we may wonder whether the decrease is reliable or due to misfits of the spectral lines. 
The modeling of the H$\alpha$ and [NII] doublet, at the nucleus, is not well constrained because of heavy line blending. Although visual inspection of the fits suggests that this assumption produces a satisfactory fit of the lines, it is possible that the amplitude of the narrow H$\alpha$ is slightly overestimated producing a decrease in the [NII]/H$\alpha$ ratio. 
Another possibility is that the decrease in the line ratios is due to an increase of the ionization parameter\footnote{That is the ratio between the density of ionizing photons and the density of the gas \citep[e.g.][]{OsterbrockF06_book}.} in the immediate vicinity of the AGN \citep[e.g.][]{FerlandNetzer83}. The decrease of the [NII]/H$\alpha$ ratio at the nucleus might be less pronounced than is the case for the other line ratios because [NII]/H$\alpha$ is more sensitive to AGN photoionization, which causes [NII]/H$\alpha$ to increase \citep{KewleyEtAl2006}.

\begin{figure}
\includegraphics[trim=0.3cm 0cm 0.3cm 0.5cm, clip=true, scale=0.63]{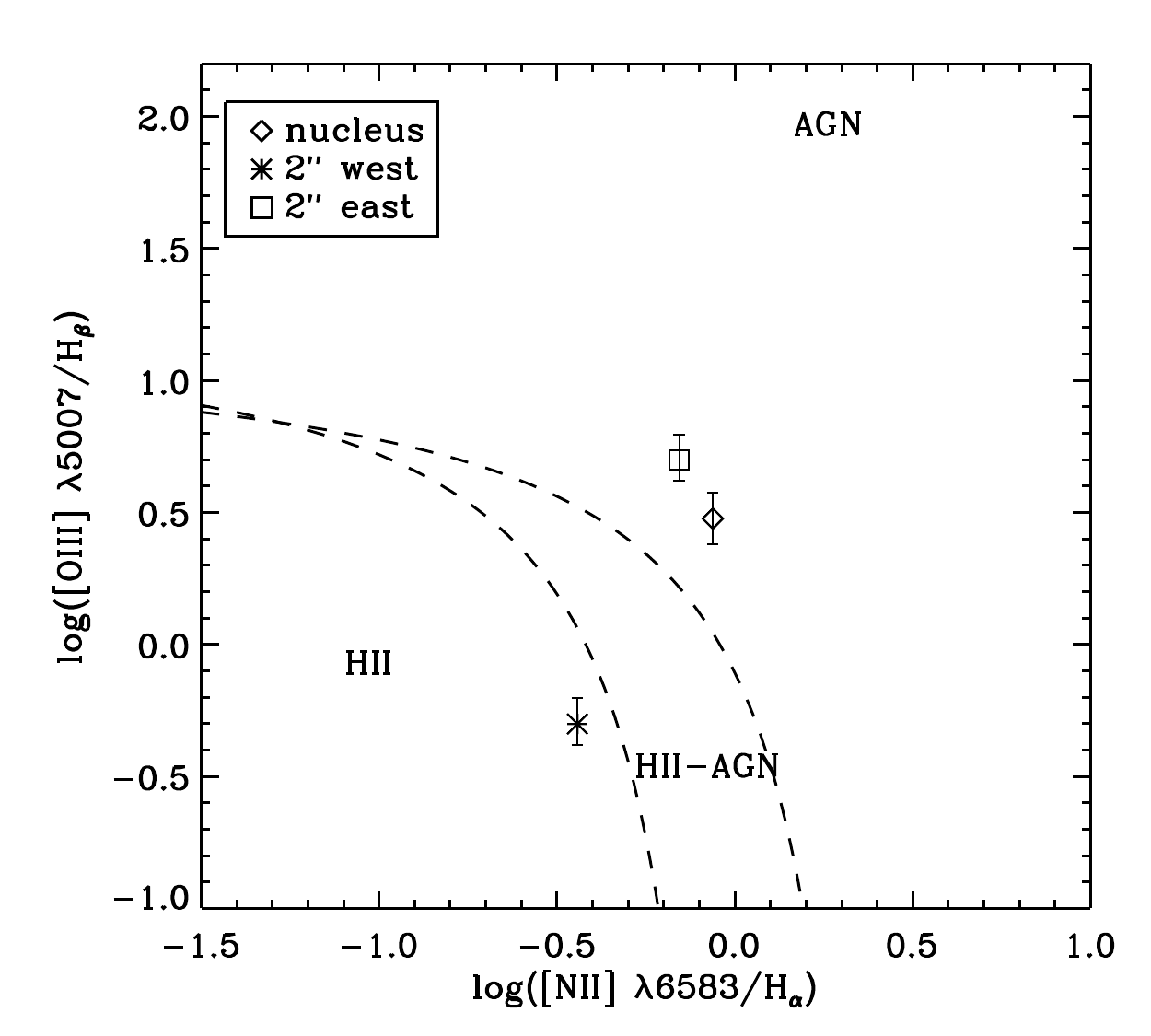}\\
\includegraphics[trim=0.3cm 0cm 0.3cm 0cm, clip=true, scale=0.63]{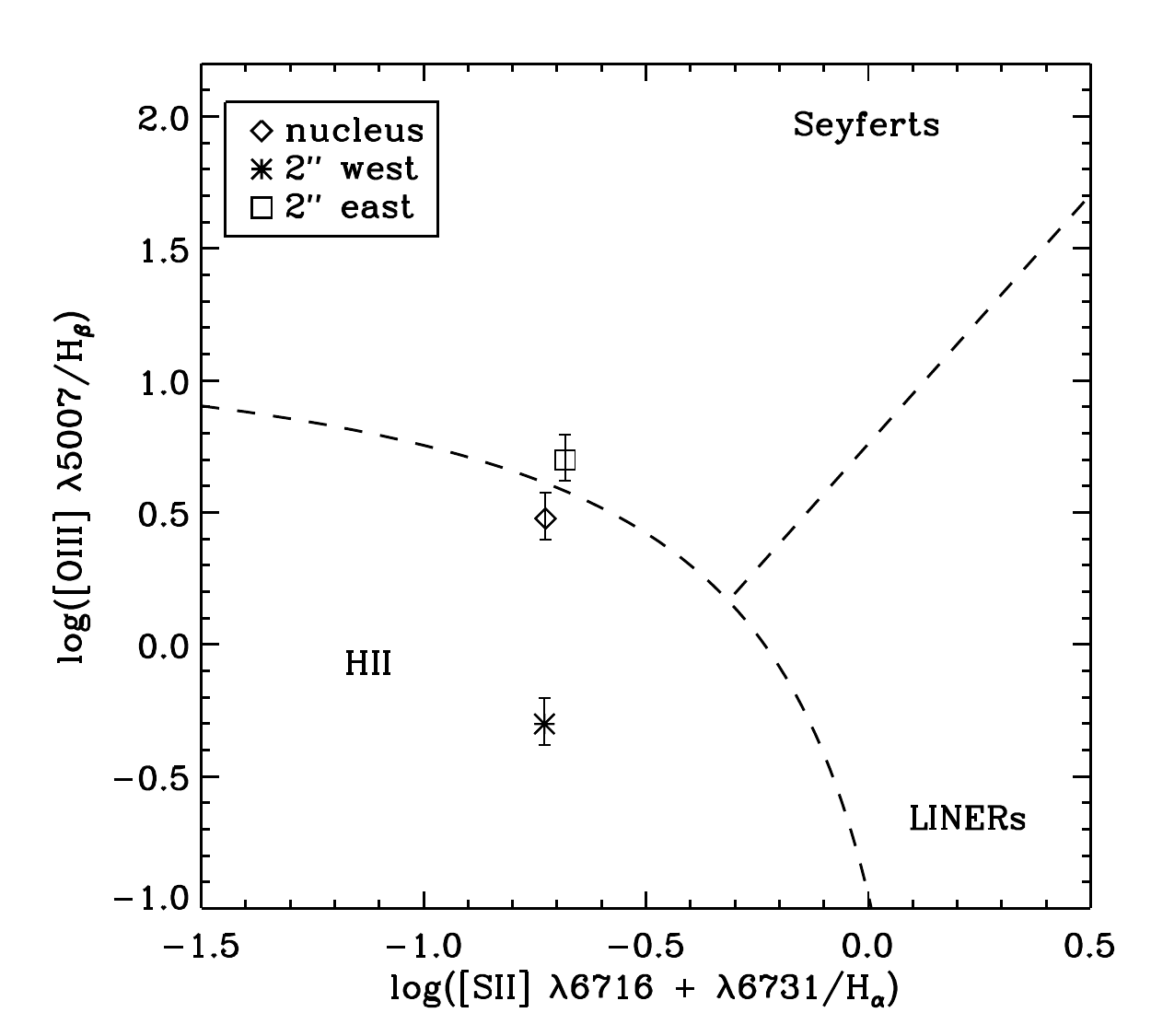}\\
\includegraphics[trim=0.3cm 0cm 0.3cm 0cm, clip=true, scale=0.63]{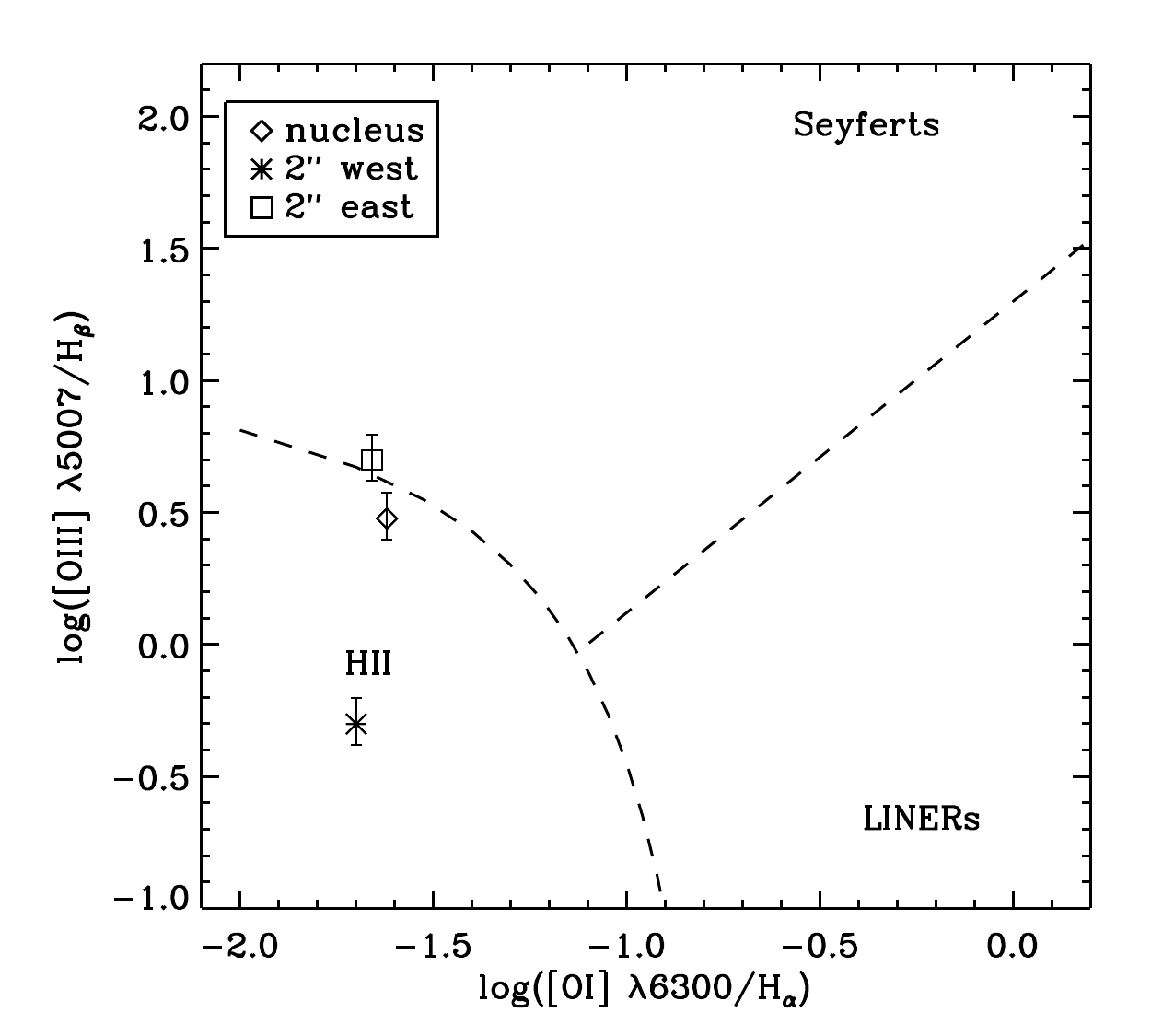}\\
\caption[BPT diagrams]{BPT diagrams for line ratios computed over circular regions of 0\farcs3 in radius centered at the nucleus, and 2$^{\prime\prime}$ east and west of the nucleus as indicated in the middle panel of Fig.\ref{fig: ratios}. Values for the [OIII]/H$\beta$ ratio have been taken from \cite{SchulzKSM99}. Horizontal error bars are approximately of the same size of the symbols. The dashed boundary lines are taken from \citet{KewleyEtAl2006}.}
\label{fig: BPT}
\end{figure}

\citet{SchulzKSM99} provide values for the [OIII]/H$\beta$ ratio at three locations within our FOV: at the nucleus, 2$^{\prime\prime}$ east and 2$^{\prime\prime}$ west of the nucleus. We derived [NII]/H$\alpha$, [SII]/H$\alpha$ and [OI]/H$\alpha$ line ratios at those locations within an extraction aperture of radius 0\farcs3 (see the middle panel in Fig.\ref{fig: ratios}). The corresponding values are plotted in the BPT diagrams \citep{BPT81} presented in Fig.\ref{fig: BPT}.

Bearing in mind that [OIII] and H$\beta$ line fluxes were not obtained simultaneously with the remaining line fluxes, nor at the same spatial resolution, it is worth noting that one point (``west") falls in the HII region of all three diagrams, whereas the other two (``nucleus" and ``east") fall close to the boundary with the AGN region. This suggests that AGN and stellar photoionization mechanisms coexist with similar energetics even at the very nucleus of the galaxy.


\section{Discussion} \label{sec: disc}

\subsection{Non-circular motions in the gas velocity field}
\label{subsec: gasv_model}

With the goal of isolating non-circular motions, we fitted the [NII] velocity field with a kinematic model describing circular orbits in a plane \citep{VDKruitAllen78, BertolaBDSS91}: 

\begin{align} \nonumber
\label{eq: vmod}
&v_\mathrm{mod}(r,\psi) 	 = v_\mathrm{sys} +\\ 
		& \frac{Ar\ \mathrm{cos}(\psi - \psi_{0})\ \mathrm{sin}\theta \ \mathrm{cos}^{p}\theta}{\{r^{2}[\mathrm{sin}^{2}(\psi - \psi_{0}) + \mathrm{cos}^{2}\theta\ \mathrm{cos}^{2}(\psi - \psi_{0})] + c^{2}_{0}\mathrm{cos}^{2}\theta\}^{p/2}}.
\end{align}

\noindent This yields a velocity curve that increases linearly at small radii and becomes proportional to $r^{(1-p)}$ at large radii.
The parameter $v_\mathrm{sys}$ is the systemic velocity, $A$ is the amplitude of the rotation curve, $r$ and $\psi$ are the radial and angular coordinates of a given pixel in the plane of the sky, $\psi_{0}$ is the position angle of the line of nodes measured with respect to the image $x$-axis (increasing counter-clockwise). The disk inclination is $\theta$ ($\theta$ = 0 for face-on disks). The parameter $p$ measures the slope of the rotation curve where it flattens, in the outer region of the galaxy, varying in the range 1-1.5. Finally, $c_{0}$ is a concentration parameter which gives the radius at which the velocity reaches 70\% of the amplitude $A$. 

As shown in \citet{ZanmarSSW08} and other previous studies, the velocity curve of NGC 1365 is still rising within the region probed by our observations, therefore the parameters $p$, $A$ (and therefore $c_{0}$) are poorly constrained by the data. We assume $p=1$, which corresponds to an asymptotically flat rotation curve at large radii. We keep $A$ and $c_{0}$ as free parameters, however these should not be taken as characteristic of the large-scale rotation curve of the galaxy. The model fitted here has the sole purpose of isolating non-circular motions.

The inclination $\theta$ has been estimated in previous works from both photometry ($\theta = 51^{\circ}$, e.g. \citealt{Lindblad78,ZanmarSSW08}) and the large-scale kinematics ($\theta = 41^{\circ}$, e.g. \citealt{JorsaterM95,ZanmarSSW08}). We adopted the latter value, $\theta = 41^{\circ}$, holding it fixed in the fit. Both \citeauthor{JorsaterM95} and \citeauthor{ZanmarSSW08} show that the photometric value produces a poor fit to the kinematics, suggesting that the discrepancy is likely due to the strong spiral features located close to the major axis \citep[e.g.][]{BarnesS03}.

\begin{figure}
\includegraphics[trim=0.7cm 1.8cm 0cm 0.5cm, clip=true, scale=0.51]{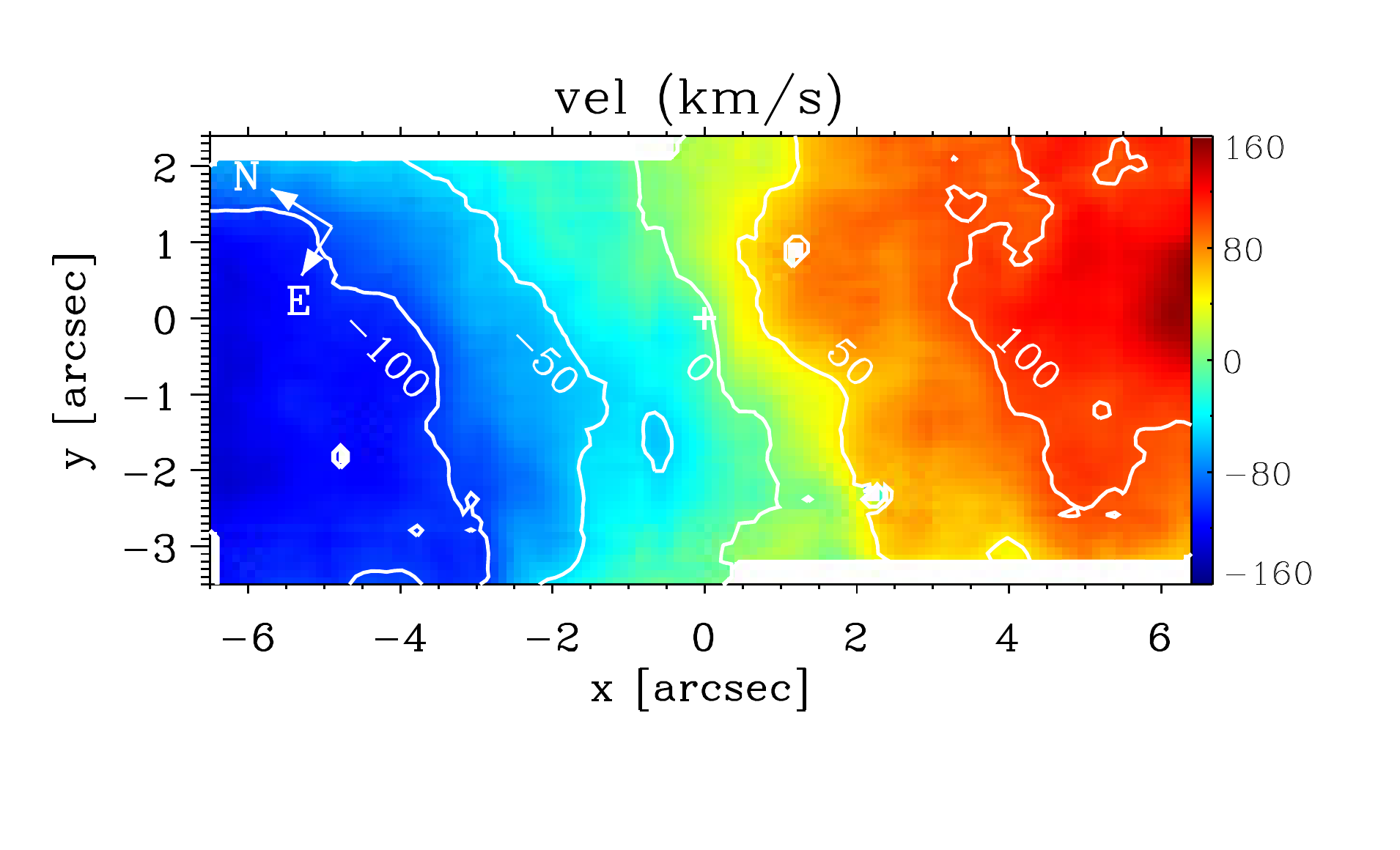}\\
\includegraphics[trim=0.7cm 1.8cm 0cm 0.5cm, clip=true, scale=0.51]{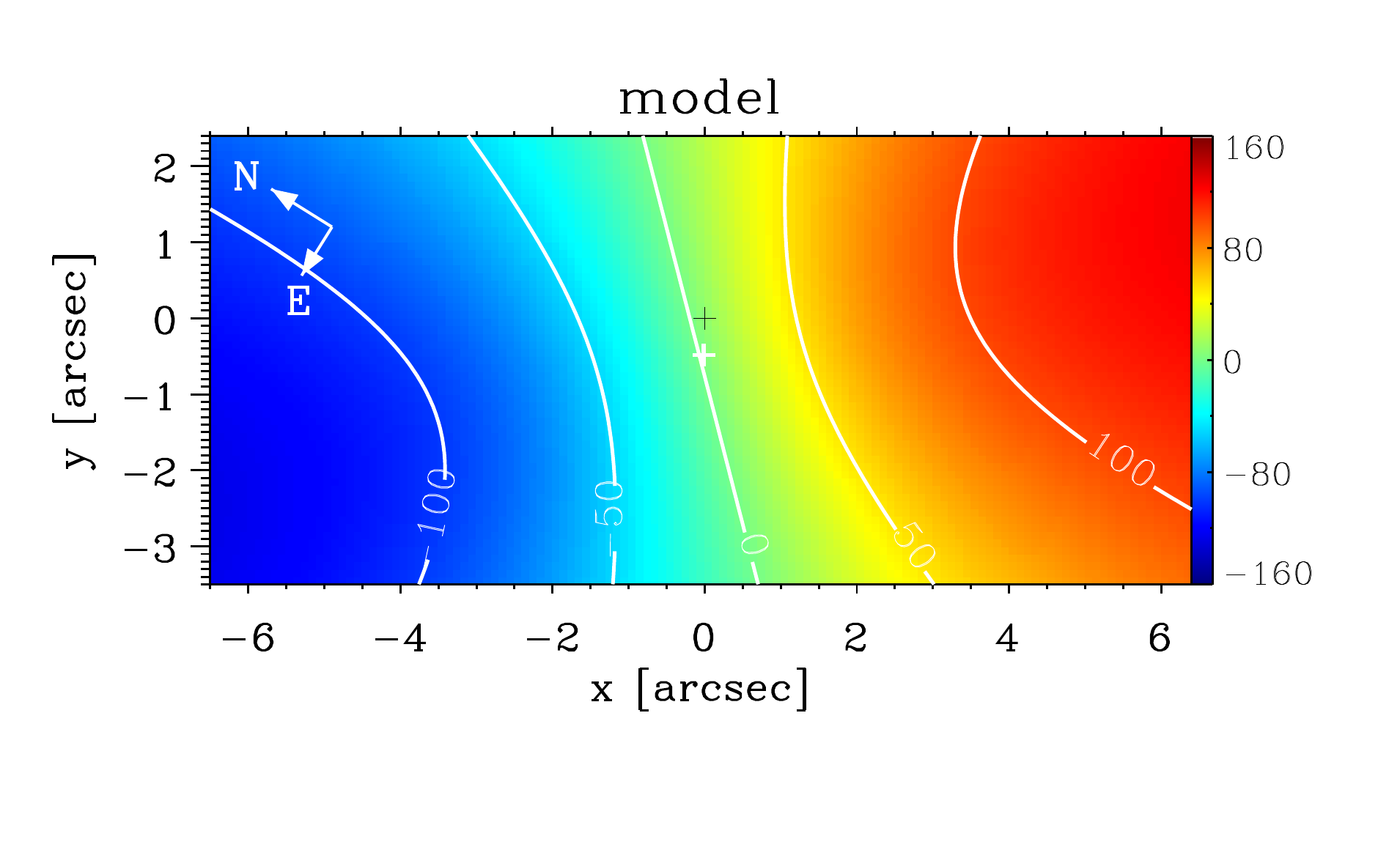}\\
\includegraphics[trim=0.7cm 1.8cm 0cm 0.5cm, clip=true, scale=0.51]{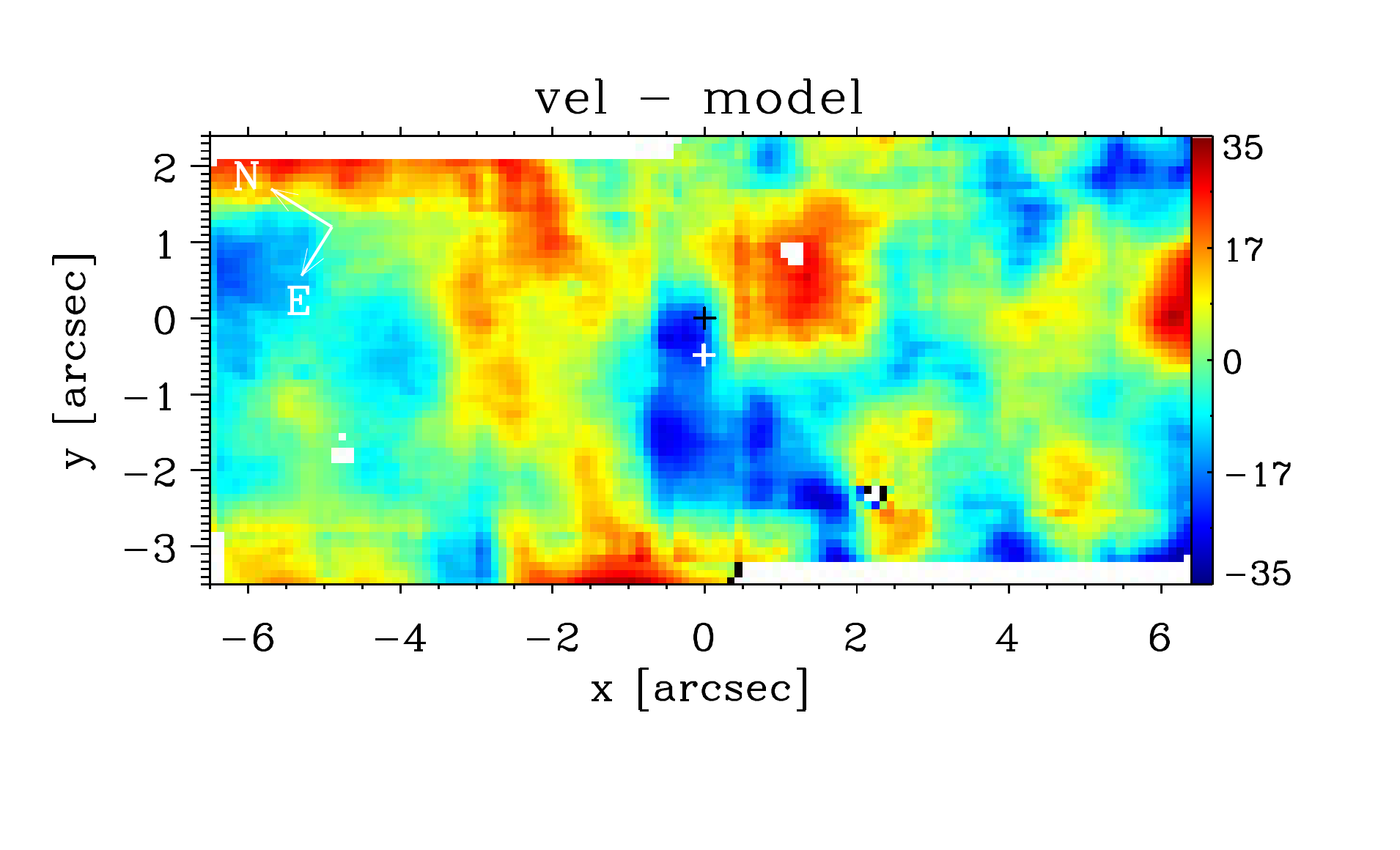}
\caption{Maps derived from the fit to the narrow component of the [NII] emission line. \textit{Top}: observed velocity field. \textit{Center}: model. \textit{Bottom}: residual. The velocity map is shown after subtracting a systemic velocity of 1671 km s$^{-1}$, derived from the fit. The black cross at (0,0) marks the continuum brightness peak. The white cross marks the kinematical center derived from the fit.} \label{fig: rotation_model}
\end{figure}

To determine the parameters ($A$, $v_\mathrm{sys}$, $\psi_{0}$, c$_{0}$) and the center of the rotation field ($x_{0}$, $y_{0}$), we fitted the rotation model defined by eq.\ref{eq: vmod} to the [NII] velocity map. The fit was performed with a customized IDL routine which makes use of the fitting engine \textit{mpfit} \citep{Markwardt09} to implement a Levenberg-Marquardt least-squares algorithm. 
The derived parameters are listed in Table \ref{tab: vmod_gas_totl}. 
Maps of the observed and the best fit model velocity field are shown in Fig.\ref{fig: rotation_model} together with the residual map.

The position angle of the line of nodes that we derived from the fit (250$^{\circ}$ east of north) is substantially different from the value derived from the large-scale kinematics (220$^{\circ}$, \citealt{JorsaterM95,ZanmarSSW08}). This is not surprising because the galaxy is strongly barred and we probed a much smaller region than that studied by \citeauthor{JorsaterM95} and \citeauthor{ZanmarSSW08}. However, a visual comparison between the [NII] velocity map derived here and the large-scale H$\alpha$ velocity map presented in Fig.4 of \citeauthor{ZanmarSSW08} shows that we obtain consistent results for the gas kinematics in the inner 6$^{\prime\prime}$.

The systemic velocity that we obtained from the fit of the kinematical model is 1671 $\pm$ 6 km s$^{-1}$ (heliocentric). This value is significantly different from the value derived from the large-scale gas kinematics \citep[1632 km s$^{-1}$,][]{JorsaterM95,ZanmarSSW08}. However, it is consistent with the average systemic velocity derived from the optical measurements listed on the Hyperleda extragalactic database\footnote{http://leda.univ-lyon1.fr/}, that is 1657 $\pm$ 10 km s$^{-1}$.

We found that the kinematical center is offset 0\farcs5 south-east of the continuum peak (comparable with the PSF FWHM, which is approximately 0\farcs6). From the large-scale gas kinematics \citeauthor{ZanmarSSW08} found an offset of a few arcseconds.

The velocity residuals have an amplitude distribution that is well represented by a Gaussian approximately centered at 1 km s$^{-1}$ with a FWHM of 27 km s$^{-1}$. Interpretation of the residual spatial distribution is not straightforward: the bottom panel in Fig. \ref{fig: rotation_model} shows that a number of deviations from circular motion are present over the field. Most of such deviations take the form of blobs with a typical size of about 1\arcsec\ and an amplitude in the range 10 - 20 km s$^{-1}$. There is no clear correspondence between the residuals and the spatial flux distribution of the H$\alpha$ emission line. In particular, the bright knots NE and SW of the nucleus do not correspond to features in the velocity residual map. Therefore, the residuals do not appear to be associated with the circum-nuclear star forming regions. 

The most interesting feature is a negative residual extending south-eastward from the continuum peak down to the edge of the FOV. This extended blueshifted residual derived from the [NII] velocity field 
is located within the well known [OIII] ionization cone \citep[e.g.][]{StorchiBergmannB91,SharpB10}.

\subsubsection{Uncertainties}

Uncertainties due to the noise on the parameters presented in Table \ref{tab: vmod_gas_totl} are negligible. 

We tested the dependence of the recovered parameters on the initial guess and we found that, using initial values drawn randomly from the range specified in Table \ref{tab: vmod_gas_totl}, the solution was stable converging, almost always, to the same values. In only a few cases (5/200) did the algorithm recover a solution which was clearly not acceptable. It seems, therefore, that the solution is robust against a fairly large range of initial guesses.

The main source of uncertainty seems to be the wavelength calibration. 
As already stated in \textsection \ref{sec: analysis}, the random error was estimated to be 6 km s$^{-1}$. However, the calibration can be affected by systematics which are difficult to assess because of the lack of suitable sky lines within the observed spectral range. 
To understand how a systematic error on the wavelength calibration would affect our results, we repeated the modeling of the velocity map after adding (subtracting) a constant value of 40 km/s (which is the difference between the systemic velocity measured here and the value derived by \citealt{JorsaterM95} and \citealt{ZanmarSSW08}). We found that the pattern and amplitude of the velocity residuals is virtually unchanged. Apart from the systemic velocity, for which we measure differences of 40 km/s, as expected, none of the other parameters shows any significant variation.


\begin{table}
\caption{FITS TO THE OBSERVED GASEOUS VELOCITY FIELD}  
\begin{tabular}{l lll}
\hline
\hline
	 Parameter				&  						& Notes					& Initial guess	\\
\hline
 & & \\
 	A [km s$^{-1}$] ........		&  218					& $\ldots$						& 100:400  \\
	$v_\mathrm{sys}$ [km s$^{-1}$] ....	& 1671 $\pm$ 6	& heliocentric, $^{\star}$			& 800:2400 \\
	$\psi_{0}$ [deg] ............. 		& 15						& $^{\dagger}$	 				& 10:60$^{\circ}$ \\
	c$_{0}$ [arcsec] .........		& 3.6						& $\ldots$						&  0:10\\
	p .......................			& 1						& fixed 						& $\ldots$\\
	$\theta$ [deg] ...............		& 41						& fixed						& $\ldots$\\
	x$_{0}$ [arcsec] .........		& 0						& $\ldots$						& -2:2\\
	y$_{0}$ [arcsec] .........		& -0.5					& $\ldots$						& -2:2\\
 & & \\	

\hline
\end{tabular}
\label{tab: vmod_gas_totl}
Notes -- $^{\star}$ The systemic velocity derived from the narrow lines at the continuum peak is 1659 $\pm$ 6 km s$^{-1}$. $^{\dagger}$ This value corresponds to a position angle on the sky of 250$^{\circ}$ east of north.
\end{table}

\subsection{Evidence for an outflow?}

Since the early observations of NGC 1365 \citep[e.g.][]{BurbidgeB60,BurbidgeBP62} it has been clear that the circum-nuclear gas kinematics is complex and that outflows could be present. Two decades later \citet{PhillipsEPT83} observed split [OIII] lines a few arcseconds from the nucleus and, inspired by the work that \citet{AxonT78} had recently published on M82, they proposed a model where the [OIII] emission originated from material outflowing along the walls of a hollow cone. A one-sided fan-shaped morphology was observed in [OIII] by, e.g., \citet{EdmundsTT88} and \citet{StorchiBergmannB91}, while later observations provided evidence for a counter fan \citep{VeilleuxSR03,SharpB10}.

\begin{figure}
\includegraphics[trim = 0cm -1cm 0cm 0.525cm, clip = true, scale = 0.33]{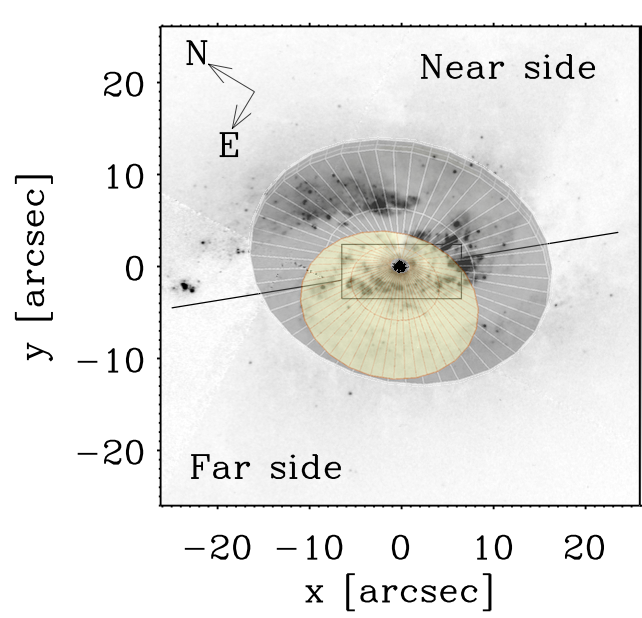}\\
\includegraphics[trim=0.5cm 1.8cm 2.885cm 0.5cm, clip=true, scale=0.51]{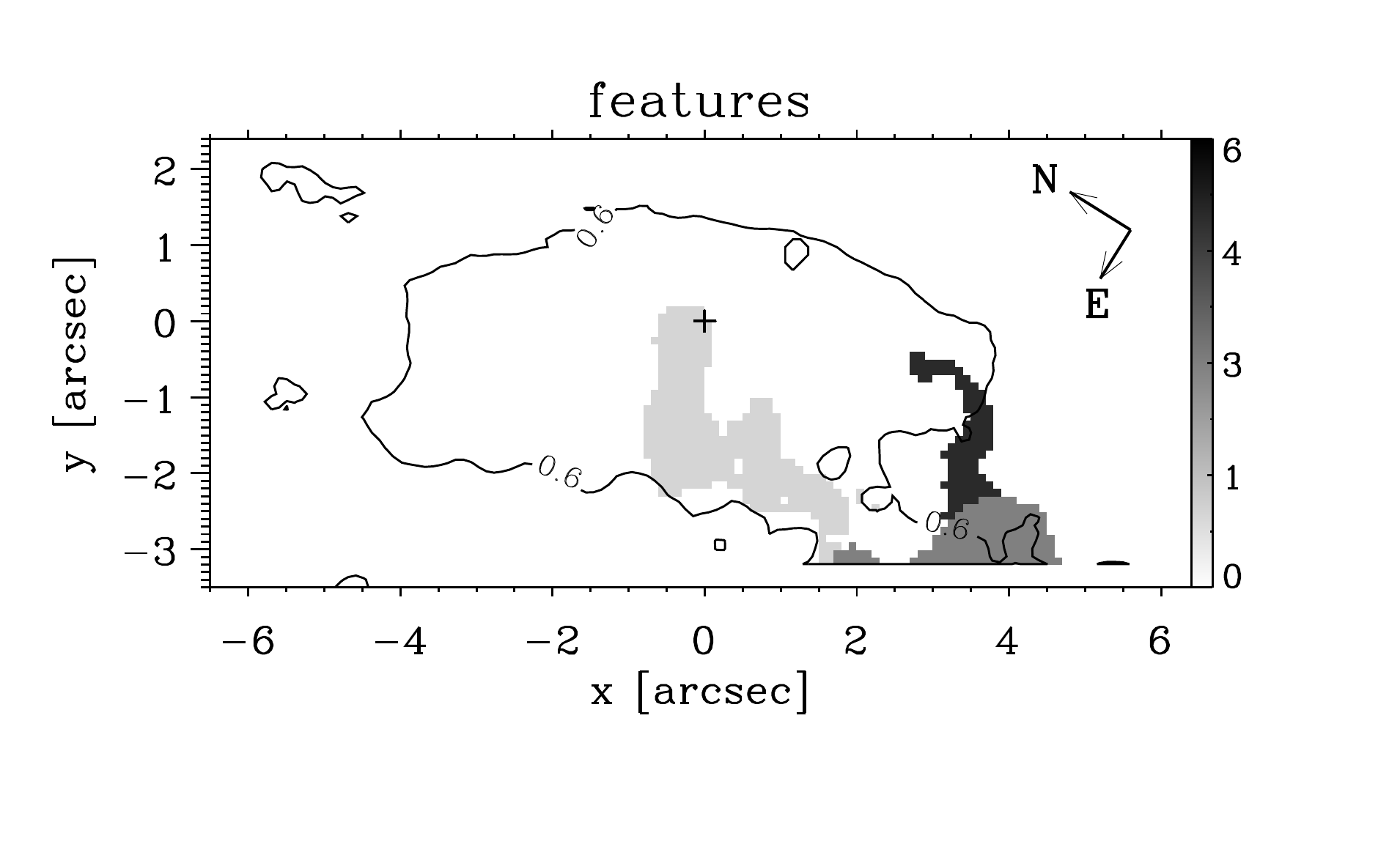}
\caption{Top: as in Fig.\ref{fig:fov}, with a cartoon representing the plane of the galaxy (large grey disk) and the approaching AGN radiation cone (in yellow, cartoon made with the \textit{Shape} software, \citealt{Steffen2010}). Bottom: map of the observed features suggestive of outflowing gas: fan-shaped blueshifted velocity excess coincident with enhanced electron density (light gray), clear evidence for a second spectral component, suggestive of unresolved split lines (dark grey), flat-topped line profiles suggestive of unresolved split lines (black and dark grey). The solid line corresponds to [NII]/H$\alpha$ = 0.6, as in Fig.\ref{fig: ratios}. Points inside the contour are characterized by [NII]/H$\alpha$ $\geq$ 0.6, suggestive of AGN photoionization.}
\label{fig: features}
\end{figure}

A number of observations were performed to confirm and understand the line splitting, and a number of models were proposed \citep[e.g.][]{EdmundsTT88,JorsaterLB84,HjelmL96}. The current consensus is that the velocity field derived from low-ionization lines (e.g. H$\alpha$ and [NII]$\lambda$6585) is consistent with rotation in a disk, while the velocity field derived from [OIII] traces a biconical outflow. According to the model proposed in \citet{HjelmL96}, the line splitting is due to the superposition of two spectral components: one originating from gas rotating in the galaxy disk, and one outflowing in a cone. The cone has a half opening angle of 50$^{\circ}$, the axis is within 5$^{\circ}$ of the galaxy rotation axis, and the outflow is radially accelerated with higher velocities close to the cone axis. See the top panel in Fig.\ref{fig: features}, or \citet{HjelmL96} and \citet{Lindblad99}, for illustrations of the geometry described above.

Evidence for an outflow coincident with the extended [OIII] emission is not obvious in the velocity field derived from the [NII] emission. However, as described in the previous section, the velocity residual map (Fig.\ref{fig: rotation_model}) shows an extended, blueshifted residual with an amplitude reaching $-30$ km s$^{-1}$, which is localized within the conical [OIII] emission mapped in previous works. While it is likely that this residual is associated with the approaching side of the AGN bipolar outflow, it is possible that the adjacent redshifted residual, approximately centered at (1$^{\prime\prime}$, 0\farcs5), is due to the receding counterpart. 

The BPT diagrams shown in Fig.\ref{fig: BPT} and the line ratio maps in Fig.\ref{fig: ratios} suggest that the residuals discussed above are located within, or adjacent to, a region dominated by AGN photoionization. However, to reach a definitive conclusion it is necessary to obtain spatially resolved, single-epoch observations of all emission lines used in the BPT diagrams (the [OIII] and H$\beta$ emission lines were not covered by our observation). The residual could be due to unresolved spectral features generated by the outflowing gas. 

As already discussed in \textsection \ref{sec: fitting}, visual inspection of the spectra revealed the presence of a line asymmetry, which we mapped by fitting GH polynomials to the emission lines, see Fig.\ref{fig: h3}. Blueshifted asymmetries are the spectral signatures expected to arise from outflowing gas moving towards the observer. It is therefore natural to ask whether they were observed in correspondence with the blueshifted residual. Unfortunately, the region where the extended blueshifted residual was identified could not be fitted with GH polynomials because of the presence of the broad H$\alpha$ emission line.

The spatial distribution of the asymmetries is somewhat puzzling: the strongest blueshifted asymmetries are present over the entire right-hand side of the FOV (south-west and south-east of the nucleus), that is, in the receding side of the galaxy (see Fig.\ref{fig:fov} or Fig.\ref{fig: features} to compare with near and far side of the galaxy). Conversely, mostly redshifted asymmetries, but with smaller amplitude than the blueshifted counterparts, are localized  on the left side of the FOV (north-west and north-east of the nucleus). No evidence of significant asymmetries is recovered south-east of the nucleus, where the conical [OIII] emission has been observed. 

The velocity field associated with the asymmetry is also puzzling. With the goal of disentangling the rotating gas from the gas responsible for the asymmetry, we fitted the lines with two Gaussians. The velocity map for the component associated with the asymmetry is presented in the top panel of Fig.\ref{fig: asym}. With the exception of a few pixels, the right side of the FOV (south-west of the nucleus) is characterized by positive velocities with a Gaussian distribution of mean value approximately equal to 50 km s$^{-1}$. Conversely, the left side of the FOV, where only few pixels were fitted with two Gaussians, shows typical velocities of $-25$ km s$^{-1}$. There is no obvious relation with the presence of a biconical outflow.

As described in \textsection \ref{sec: eden}, we used the [SII] doublet to derive the electron density. Remarkably, Fig.\ref{fig: density} shows that a sharp increase in the electron density is spatially coincident with the fan-shaped blueshifted residual, a typical kinematical signature of outflows. In previous works, density enhancements have been observed in coincidence with kinematical features interpreted as due to outflowing gas \citep[e.g.][]{HoltTM11,SMSBN14,LenaRSB14,VillarMartinBS2015}.
In their study of the narrow-line region of Mrk 477, \citealt{VillarMartinBS2015} argue that the enhanced electron density associated with the outflow might be due to a shock induced by the radio jet. In the case of NGC 1365, this hypothesis is unlikely: first, as discussed below, there is little evidence of a radio jet; second, as shown by the BPT diagrams in Fig.\ref{fig: BPT}, the line ratios $2^{\prime\prime}$ east of the nucleus are not characteristic of shock-induced ionization (i.e. LINER-like).

The spatial coincidence between the density enhancement and the velocity residuals suggests a connection between the two: it is conceivable that the enhancement is due to a wind of denser gas, perhaps blown off the torus, and accelerated by the AGN radiation pressure \citep[e.g. \textsection 4 in][and references therein]{CrenshawKG03MassLoss}.
\vskip10pt

In previous works, non-circular motions in NGC 1365 have been mapped via the modeling of split line profiles in [OIII] and other high-ionization species. Split lines are not resolved within our FOV, nevertheless from visual inspection of the datacube it is obvious that a second blueshifted component is present, in H$\alpha$, [NII] and [SII], in a handful of pixels near the edge of the FOV. More precisely, this region is characterized by (i) an enhancement in the velocity dispersion located around the point (4$^{\prime\prime}$, -3$^{\prime\prime}$), see middle panel in Fig.\ref{fig: vel_narrow}; (ii) by an evident decrease in the GH coefficient $h_{4}$, a proxy for the line kurtosis, indicating line profiles with a flat top, see bottom panel in Fig.\ref{fig: h3}. The same region has enhanced blueshifted velocity residuals as large as -27 km s$^{-1}$, see bottom panel in Fig.\ref{fig: rotation_model}.

In summary: features suggestive of outflowing gas have been so far observed in high-ionization species. Here we found similar evidence also in a number of low-ionization species, i.e. [NII], H$\alpha$, [SII], and [OI]: we observed blueshifted velocity residuals in a fan-shaped region with the apex at the nucleus, oriented in the same way as the [OIII] conical emission, with line ratios suggestive of AGN photoionization, and enhanced electron density; evidence for line splitting was identified in an adjacent area. A composite map of the spatial distributions of these features is presented in the bottom panel of Fig.\ref{fig: features}.

\vskip10pt
Outflows can be powered by AGNs \citep[e.g.][]{VeilleuxCB05,CrenshawKG03MassLoss}, star-formation activity \citep[e.g.][]{AxonT78,VeilleuxR02}, or a combination of the two \citep[e.g.][]{CecilBV01}. If a radio jet is present, then it can influence the morphology and kinematics of the outflowing gas \citep[e.g.][]{AxonMC98,MazzalayRK13}. 
It has been suggested that NGC 1365 hosts a radio jet \citep{SandqvistJS95}. However, observations reported by \citet{StevensFN99} show that the bulk of the radio emission in the nuclear region is due to an elongated ($8^{\prime\prime} \times 20^{\prime\prime}$) star-forming ring. The jet-like feature identified by \citeauthor{SandqvistJS95} is seen to connect the nucleus to the ring, but if it is a jet, it is small and extremely weak.
The galaxy is known to host both an AGN and intense star-formation activity. Therefore, it is natural to ask whether the outflow is predominantly powered by the AGN or by the star-forming regions. The wide-field IFU data obtained by \citet{SharpB10} suggest that the outflowing gas is ionized by the AGN. Our data, which probe smaller scales at higher spatial resolution, provide supporting evidence that the AGN is powering the outflow: namely, the base of the velocity residual - which we interpret as a tracer of the outflow - is coincident with the unresolved nucleus. Therefore, the base of the outflow falls well inside the star-forming ring. Moreover, the presence of a broad H$\alpha$ line, and the line ratios presented here, support the hypothesis that the AGN is indeed the predominant ionizing source of the outflowing gas.

\subsection{Is there any evidence for gas inflows?}
\label{subsec: asymmetry_N1365}
Large-scale bars are among those dynamical features that funnel gas towards the inner region of their host galaxies. NGC 1365 has a strong bar with a projected major axis extending over about $200^{\prime\prime}$ \citep[e.g.][]{ZanmarSSW08}. Our observation, with an FOV of $13^{\prime\prime}\times6^{\prime\prime}$ centered on the nucleus, probed only a very small fraction of the bar, in its central region. According to the hydrodynamical models of NGC 1365 presented in \citet{LindbladLA96}, this region falls entirely within the inner Lindblad resonance, which is located at a radius of $r \approx 30^{\prime\prime}$. Their model predicts the presence of spiral gas inflows within this region, however it must be acknowledged that it relies on a number of simplifying assumptions and, similarly to other studies \citep[e.g.][]{ZanmarSSW08,PinolFLF12}, it attempts to reproduce the main features of the large-scale kinematics. The spatial resolution of the model is therefore not ideal for a direct comparison with our results.

The velocity residuals derived from the [NII] emission line do not show any obvious pattern which could be associated with gas inflows. It is worth recalling that a distinctive trait of the narrow emission lines is the presence of an asymmetric base. This is systematically blueshifted with respect to the line peak in the right-hand (south-western) portion of the FOV and redshifted in the left side. 
When the emission lines are modeled with two components, the asymmetry shows velocities which are lower than the velocities derived from the strongest spectral component, i.e. the narrow core (compare the velocity fields in Fig.\ref{fig: vel_narrow} and Fig.\ref{fig: asym}). Therefore, the asymmetric base could be tracing gas which is rotating more slowly than the gas responsible for the bulk of the line emission. In other words, it could originate from gas which lost angular momentum - perhaps via shocks along the edge of the large-scale bar -  and it is slowly migrating from the inner Lindblad resonance towards the nucleus. In support of this speculation, we note that the observed region is embraced by two evident dust lanes, see top left panel in Fig.\ref{fig:fov}. These features are usually interpreted as tracers of gas which has been compressed, or shocked, at the leading edges of a bar \citep[e.g. \textsection 6.5 in][]{BTGalacticDynamics}.

\section{Summary and Conclusions} 
\label{sec: conclusion_NGC1365}
We observed the Seyfert 1.8 galaxy NGC 1365 using the GMOS integral field unit on the GEMINI South telescope. The FOV was centered on the nucleus and is aligned with the large-scale bar, covering 1173 $\times$ 541 pc$^{2}$. Our observation covered the spectral range 5600 - 7000 \AA\ which includes a number of emission lines ([OI]$\lambda$6300, [NII]$\lambda\lambda$6548,6583, H$\alpha$ and [SII]$\lambda\lambda$6716,6731). We modeled the profiles of these lines to produce velocity, velocity dispersion and flux maps with a spatial resolution of 52 pc and a spectral resolution of 49 km s$^{-1}$. 

The flux maps are dominated by the strong and compact nuclear emission, while several bright knots, probably HII regions, are present elsewhere. The bright emission at the nucleus includes a prominent and unresolved broad component in the H$\alpha$ emission line; because of the PSF wings, this contributes emission out to a radius of approximately 2$^{\prime\prime}$ from the nucleus. The broad component was modeled with a combination of two Gaussians from which we derived a combined velocity dispersion of 1181 km s$^{-1}$ and a flux-weighted velocity of -100 km s$^{-1}$ (with respect to a systemic velocity of 1671 km s$^{-1}$, as derived from the [NII]$\lambda$6583 emission line).

Strong narrow emission lines are present over the entire FOV with velocity dispersion in the range 35 - 105 km s$^{-1}$ and mean value $\sigma \approx$ 62 km s$^{-1}$. 
The velocity fields derived from the [NII]$\lambda$6583 emission line and, independently, from H$\alpha$, [SII], and [OI], are consistent with gas rotating in the large-scale disk, however extended blueshifted velocity residuals suggest the presence of an additional kinematical component. 
The residual has a fan-shaped morphology with the apex at the nucleus; it is located within a region where previous authors observed an extended [OIII] emission thought to trace a conical outflow \citep[e.g.][]{StorchiBergmannB91}; it corresponds to a sharp increase in the electron density (from 150 to 500 cm$^{-3}$), and it is characterized by line ratios typical of AGN photoionization ([NII]/H$\alpha$$>$0.6). Close to this region we find evidence for line splitting. 
These features, previously observed at larger radii and for higher ionization species, are consistent with the hypothesis of a conical outflow suggested in previous works \citep[e.g.][]{HjelmL96}. The presence of a broad H$\alpha$, the line ratios and the morphology of the velocity residuals support the hypothesis that the outflow is predominantly powered by the AGN.

Finally, although NGC 1365 shows overall morphological similarities with NGC 1097, we found no obvious evidence for gas inflows along nuclear spirals, like those observed in NGC 1097 by \citet{FathiSBREtAl06}. However, the emission lines show a weak but evident asymmetry which is blueshifted with respect to the core of the lines in the right-hand side of the FOV, and redshifted in the left-hand side. This asymmetry could arise from gas which is rotating more slowly than the gas responsible for the bulk of the narrow line emission. We speculate that it could be tracing gas which lost angular momentum, perhaps via the shocks traced by the adjacent prominent dust lanes, and it is now in the process of slowly migrating from the inner Lindblad resonance towards the nucleus of the galaxy.

\section*{Acknowledgements}
We thank the anonymous referee for insightful comments,
which helped to improve the manuscript. D.L. acknowledges support from the 
National Science Foundation under grant no. AST - 1108786 (PI Robinson), and from the European Research Council (ERC) under grant 647208 (PI Jonker). R.A.R. acknowledges support from FAPERGS (project no. 2366-2551/14-0) and CNPq (project no. 470090/2013-8 and 302683/2013-5).

The authors wish to recognize and acknowledge the cultural role and reverence that the summit of Mauna Kea has always had within the indigenous Hawaiian community. We are most fortunate to have the opportunity to obtain data from observations conducted from this mountain.
This work is based on observations obtained
at the Gemini Observatory, which is operated by the Association
of Universities for Research in Astronomy, Inc., under a cooperative
agreement with the NSF on behalf of the Gemini partnership:
the National Science Foundation (United States), the Science
and Technology Facilities Council (United Kingdom), the National
Research Council (Canada), CONICYT (Chile), the Australian Research
Council (Australia), Ministerio da Ciencia e Tecnologia
(Brazil) and south-east CYT (Argentina).

We acknowledge the usage of the HyperLeda
database (http://leda.univ-lyon1.fr) and the NASA/IPAC Extragalactic Database (NED) which is operated by the
Jet Propulsion Laboratory, California Institute of Technology, under contract with the
National Aeronautics and Space Administration.




\bibliographystyle{mnras}
\bibliography{biblio}


\bsp	
\label{lastpage}
\end{document}